\documentclass[reprint,twocolumn,aps,pra,amsmath,amssymb,superscriptaddress,longbibliography]{revtex4-2}

\usepackage{graphicx}
\usepackage{color}
\usepackage[colorlinks,urlcolor=cyan,citecolor=blue,linkcolor=magenta]{hyperref}
\makeatletter
\newcommand*{\rom}[1]{\expandafter\@slowromancap\romannumeral #1@}
\makeatother

\begin{document}
\title{Superfluidity of a Raman spin-orbit-coupled Bose gas at finite temperature}

\author{Xiao-Long Chen}
\email{xiaolongchen@swin.edu.au}
\affiliation{Department of Physics, Zhejiang Sci-Tech University, Hangzhou 310018, China}
\affiliation{Institute for Advanced Study, Tsinghua University, Beijing 100084, China}
\affiliation{Centre for Quantum Technology Theory, Swinburne University of Technology, Melbourne 3122, Australia}

\author{Xia-Ji Liu}
\affiliation{Centre for Quantum Technology Theory, Swinburne University of Technology, Melbourne 3122, Australia}

\author{Hui Hu}
\affiliation{Centre for Quantum Technology Theory, Swinburne University of Technology, Melbourne 3122, Australia}
\date{\today}

\begin{abstract}
We investigate the superfluidity of a three-dimensional weakly interacting Bose gas with a one-dimensional Raman-type spin-orbit coupling at both zero and finite temperatures. 
Using the imaginary-time Green's function within the Bogoliubov approximation, we explicitly derive analytic expressions of the current-current response functions in the plane-wave and zero-momentum phases, from which  we extract the superfluid density in the limits of long wavelength and zero frequency. At zero temperature, we check that the resultant superfluid density agrees exactly with our previous analytic prediction obtained from a phase-twist approach. Both results also satisfy a generalized Josephson relation in the presence of spin-orbit coupling. At finite temperature, we find a significant non-monotonic temperature dependence of superfluid density near the transition from the plane-wave phase to the zero-momentum phase. We show that this non-trivial behavior might be understood from the sound velocity, which has a similar temperature dependence. The non-monotonic temperature dependence is also shared by Landau critical velocity, above which the spin-orbit-coupled Bose gas loses its superfluidity. Our results would be useful for further theoretical and experimental studies of superfluidity in exotic spin-orbit coupled quantum gases.
\end{abstract}

\maketitle


\section{Introduction}

Spin-orbit coupling (SOC), linking a particle's spin to its motion, can lead to ubiquitous quantum effects in various areas of physics~\cite{Galitski2013,zhai2015degenerate}. It plays a central role in the emergent exotic bosonic and fermionic states of matter~\cite{wang2010spin,ho2011bose,li2012quantum,hu2012spin,gong2011bcs,hu2011probing,yu2011spin}, such as topological insulators featuring quantum spin Hall effect~\cite{zhu2006spin,liu2007optically,beeler2013spin}, and topological Fermi superfluids and Majorana fermions~\cite{sato2009non,jiang2011majorana,liu2012probing,liu2012topological,*liu2013topological,liu2014realization}. Owing to its versatility~\cite{bloch2008many}, ultracold atomic gases have became a powerful platform to create synthetic gauge fields and to simulate SOC~\cite{liu2009effect,spielman2009raman,dalibard2011colloquium,goldman2014light}. In the last decade, by utilizing Raman laser beams to carefully modulate atom-light interactions, experimentalists have successfully realized one-dimensional Raman-laser-induced SOC~\cite{lin2011spin,wang2012spin,cheuk2012spin} and two-dimensional SOC~\cite{wu2016realization,huang2016experimental,chen2018spin,zhang2019ground} in both ultracold Bose and Fermi gases. These crucial achievements have driven tremendous investigations on the nontrivial role of SOC in quantum many-body physics~\cite{li2012sum,martone2012anisotropic,zheng2013properties,chen2017quantum,chen2018quantum,zhang2012collective,khamehchi2014measurement,ji2015softening,ji2014experimental}.

Superfluidity is a prominent phenomenon in quantum liquids, supporting frictionless flow in a capillary without dissipating energy~\cite{lifshitz2013statistical,pitaevskii2016bose}. This property is closely related to Bose-Einstein condensation and has been qualitatively understood by using the celebrated Landau criterion. For an isotropic system that is invariant under Galilean transformation, a critical velocity $v_c\equiv \mathrm{min}\left[\omega({\bf p})/|{\bf p}|\right]$ can be formally defined with the elementary excitation spectrum $\omega({\bf p})$ of the superfluid. Below the critical velocity, an impurity's motion can not create excitations to induce dissipation in energy and hence to destroy the superfluidity of the system. In conventional weakly-interacting Bose gases, this critical velocity is dominated by the linear phonon mode $\omega({\bf p})= c_s {\bf p}$ of the lowest-lying excitation spectrum in the long-wavelength limit, so it coincides with the sound velocity $c_s$.

SOC is known to violates the Galilean invariance~\cite{zhu2012exotic,pitaevskii2016bose}. Therefore, it becomes an intriguing question to characterize the superfluidity of a weakly-interacting Bose gas in the presence of, e.g., the Raman-induced SOC. Previous works have shown that, this Raman-type SOC can not only lead to various ground-state phases by tuning the Raman coupling (i.e., the Rabi frequency)~\cite{li2012sum,martone2012anisotropic,zheng2013properties}, i.e., the stripe (ST), plane-wave (PW), and zero-momentum (ZM) phases, but also significantly changes the excitation spectrum with the appearance of a roton-maxon structure in the PW phase and a quadratic dispersion at the PW-ZM transition~\cite{martone2012anisotropic,zheng2013properties,chen2017quantum}. As a result, the critical velocity is determined by the roton minimum instead of the phonon mode, leading to anisotropic superfluidity. The critical velocity can even vanish at the critical PW-ZM transition point~\cite{zheng2013properties,yu2017landau}. These novel properties in the elementary excitation spectrum as well as the sound velocity have already been demonstrated in cold-atom laboratories~\cite{zhang2012collective,khamehchi2014measurement,ji2015softening}.

However, there are only a few works on the calculation of superfluid density, a key quantity that directly characterizes the superfluidity of the system. The superfluid density of a SOC Bose gas was first predicted for the PW and ZM phases in Ref.~\cite{zhang2016superfluid} at \emph{zero} temperature by Zhang \textit{et al.}, using a sum-rule analysis of current-current correlation functions. The study shows that the gapped branch in the elementary excitation spectrum plays a crucial role in inducing a nonzero normal density. Most recently, by employing a first-order stripe {\it Ansatz} and a plane-wave {\it Ansatz} within a phase-twist method, the present authors and collaborators derived analytic expressions of zero-temperature superfluid density as a function of Rabi frequency in different phases~\cite{chen2018quantum}, consistent with the findings in Ref.~\cite{zhang2016superfluid}. In addition, the dependence of superfluid density on the high-order harmonics has also been investigated numerically in the exotic stripe phase. The high-order harmonics shift the critical Rabi frequency of the ST-PW phase transition point and suppress the superfluid density, compared to the results obtained by using a first-order stripe {\it Ansatz}. Recently, by means of a perturbative approach at small Rabi frequency, Martone and Stringari also addressed the superfluid density in the stripe phase at zero temperature within the sum rule approach~\cite{martone2021supersolid}.

In this work, we aim to investigate the effect of thermal fluctuations on the superfluidity of a three-dimensional weakly interacting Bose gas with Raman-type SOC, by conducting a microscopic calculation of superfluid density using current-current correlation functions at both zero temperature and \emph{finite} temperature. This direct calculation is crucial, since the previous sum-rule approach~\cite{zhang2016superfluid} is no longer applicable at finite temperature. In more detail, by means of the Gross-Pitaevskii theory within the Bogoliubov approximation, we first calculate the condensate wave function from a plane-wave {\it Ansatz} and the Bogoliubov quasiparticle wave functions with respect to the tunable Rabi frequency. The current-current response function is then derived analytically using the imaginary-time Green's function formalism. The total and normal density can be extracted from the longitudinal and transverse response functions in the long-wavelength and low-frequency limits, respectively.

To check our direct calculation of the superfluid density, at zero temperature we compare the obtained superfluid density fraction with the previous analytic expression derived from a phase-twist method~\cite{chen2018quantum}. We also consider an alternative examination by using the so-called Josephson relation, which relates the superfluid density to condensate density. Josephson relation has been studied in detail in conventional spinless weakly interacting Bose gases~\cite{hohenberg1965microscopic,josephson1966relation,baym1968microscopic,holzmann2007condensate}, and has also been considered recently in the multi-component case~\cite{zhang2018generalized}. Here, we generalize the Josephson relation to the case of a SOC Bose gas, from which we determine the zero-temperature superfluid density directly via the single-particle Green's function and the condensate density.

At finite temperature, we find a non-monotonic temperature dependence of the extracted superfluid density near the transition from the plane-wave phase to the zero-momentum phase. We discuss the causes of this non-trivial behavior in superfluid density, by considering the sound velocity and Landau critical velocity at finite temperature.

The rest of the paper is organized as follows. The model Hamiltonian and the theoretical framework are introduced in Sec.~\ref{sec:theory}. In Sec.~\ref{sec:results}, we derive the analytic expression of the current-current response functions, and show how to calculate the superfluid density via the longitudinal and transverse components of the response functions (see Fig.~\ref{fig1}). We then apply the generalized Josephson relation to spin-orbit coupled Bose gases to obtain the superfluid density. In the plane-wave and zero-momentum phases, the predictions of superfluid density at zero temperature from two approaches are compared with our previous analytic result obtained from a phase-twist method (see Fig.~\ref{fig2}). We further calculate the current-current response functions at finite temperature and study the temperature dependence of superfluid density. To understand the non-trivial temperature dependence, we also show the sound velocity and Landau critical velocity as a function of temperature (see Figs.~\ref{fig3} and~\ref{fig4}). A summary and outlook are given in Sec.~\ref{sec:summary}. Appendix A is devoted to the technical details of the current-current response functions.

\section{Theoretical Frameworks\label{sec:theory}}

\subsection{The model Hamiltonian \label{sec:model}}

We consider a three-dimensional weakly interacting spin-1/2 Bose gas with a one-dimensional Raman-induced spin-orbit coupling, the same as in our previous works~\cite{chen2017quantum,chen2018quantum}. The system can be described by the model Hamiltonian, $\hat{H}=\hat{H}_{0}+\hat{H}_{\mathrm{int}}$, where the single-particle Hamiltonian $\hat{H}_{0}$ and the interaction Hamiltonian $\hat{H}_{\mathrm{int}}$ read, respectively ($\hbar=1$ and  the superscript “*” represents the Hermitian conjugate)~\cite{lin2011spin,li2012quantum,zheng2013properties}
\begin{subequations}
\begin{eqnarray}
\hat{H}_{0} & = & \int d^{3}{\bf r}\begin{pmatrix}
\hat{\Psi}_{\uparrow}^*({\bf r}), &\hat{\Psi}_{\downarrow}^*({\bf r})
\end{pmatrix}\mathcal{H}_{\mathrm{s}}(\hat{{\bf p}})\begin{pmatrix}
\hat{\Psi}_{\uparrow}({\bf r})\\
\hat{\Psi}_{\downarrow}({\bf r})
\end{pmatrix}, \label{eq:single-particle} \\
\hat{H}_{\mathrm{int}} & = & \int d^{3}{\bf r}\sum_{\sigma,\sigma^{\prime}=\uparrow,\downarrow}\frac{g_{\sigma\sigma^{\prime}}}{2}\hat{\Psi}_{\sigma}^*\hat{\Psi}_{\sigma^{\prime}}^*\hat{\Psi}_{\sigma^{\prime}}\hat{\Psi}_{\sigma}({\bf r}). \label{eq:interaction}
\end{eqnarray}
\end{subequations}
Here, the single-particle part $\mathcal{H}_{\mathrm{s}}(\hat{{\bf p}})$ is given by 
\begin{equation}
\mathcal{H}_{\mathrm{s}}(\hat{{\bf p}})=\frac{(\hat{{\bf p}}-k_{\mathrm{r}}\hat{{\bf e}}_{x}\sigma_{z})^{2}}{2m}+\frac{\Omega}{2}\sigma_{x}+\frac{\delta}{2}\sigma_{z},
\end{equation}
with the canonical momentum operator $\hat{{\bf p}}=-i\nabla$ and Pauli matrices $\sigma_{x,z}$. $k_{\mathrm{r}}\hat{{\bf e}}_{x}$ is the recoil momentum of the Raman lasers along the $x$ axis, with a recoil energy $E_{\mathrm{r}}=k_{\mathrm{r}}^{2}/(2m)$. It is straightforward to see that the momentum operator $\hat{{p}}_x$ is coupled to the spin via the one-dimensional physical momentum term $(\hat{{p}}_x-k_{\mathrm{r}}\sigma_{z})^2$. For simplicity, the detuning of the Raman lasers is assumed to be zero $\delta=0$. The Rabi frequency $\Omega$ can be flexibly tuned, in accord with the recent experiments~\cite{ji2014experimental,ji2015softening}. $g_{\sigma\sigma^{\prime}}=4\pi a_{\sigma\sigma^{\prime}}/m$ are interaction strengths for intra- ($\sigma=\sigma^{\prime}$) and inter-species ($\sigma\neq\sigma^{\prime}$), and $a_{\sigma\sigma^{\prime}}$ are the corresponding $s$-wave scattering lengths.

\subsection{A self-consistent approach at finite temperature \label{sec:bogoliubov}} 

In this work, we employ the quasiparticle formalism at the Bogoliubov level to describe a weakly interacting dilute Bose gas with SOC at zero and finite temperature~\cite{griffin1996conserving,dodd1998collective,buljan2005incoherent,dalfovo1999theory,pitaevskii2016bose}. Following the standard procedure as in our previous works~\cite{chen2015collective,chen2017quantum,chen2018quantum}, the Bose field operator $\hat{\Psi}_{\sigma}({\bf r},t)$ for spin component $\sigma=(\uparrow,\downarrow)$ can be rewritten as a combination of the condensate wave function $\psi_{\sigma}$ and the noncondensate fluctuation operator $\hat{\eta}_{\sigma}$ as
\begin{equation} \label{eq:newBosefield}
\hat{\Psi}_{\sigma}({\bf r},t)=\psi_{\sigma}({\bf r})+\hat{\eta}_{\sigma}({\bf r},t).
\end{equation}
Using a Bogoliubov transformation, the fluctuation operator $\hat{\eta}_{\sigma}({\bf r},t)$ and its conjugate can be expanded as
\begin{subequations} \label{eq:eta}
\begin{eqnarray}
\hat{\eta}_{\sigma} &=& \underset{j}{\sum}\left[u_{j\sigma}({\bf r})e^{-i\omega_{j}t}\hat{a}_{j}+v_{j\sigma}^{*}({\bf r})e^{i\omega_{j}t}\hat{a}_{j}^{\dagger}\right],\\
\hat{\eta}_{\sigma}^* &=& \underset{j}{\sum}\left[u_{j\sigma}^{*}({\bf r})e^{i\omega_{j}t}\hat{a}_{j}^{\dagger}+v_{j\sigma}({\bf r})e^{-i\omega_{j}t}\hat{a}_{j}\right],
\end{eqnarray}
\end{subequations}
in terms of the quasiparticle wave functions $u(u^{*})$, $v(v^{*})$ and the quasiparticle frequency $\omega_{j}$. In free space, the Bogoliubov wave functions $u$, $v$ can be expanded as $u_{j\sigma}({\bf r})=u^{(\lambda)}_{{\bf q}\sigma}e^{i{\bf qr}}/\sqrt{V}$ and $v_{j\sigma}({\bf r})=v^{(\lambda)}_{{\bf q}\sigma}e^{i{\bf qr}}/\sqrt{V}$, where $V$ is the system volume. Here, $j\equiv({\bf q},\lambda)$ is the index of the quasiparticle energy level, with the momentum ${\bf q}$ and the branch index $\lambda$. $\hat{a}^{\dagger}$ ($\hat{a}$) are respectively the creation (annihilation) operators for quasiparticles, satisfying the bosonic commutation relations $[\hat{a}_{i},\hat{a}^{\dagger}_{j} ]=\delta_{ij}$, $[\hat{a}_{i}^{\dagger},\hat{a}_{j}^{\dagger}]=[\hat{a}_{i},\hat{a}_{j}]=0$. After substituting Eqs.~\eqref{eq:newBosefield} and~\eqref{eq:eta} into the equations of motion
\begin{equation} \label{eq:eq-of-motion}
i\partial_{t}\hat{\Psi}_{\sigma}({\bf r},t) = \begin{bmatrix} \hat{\Psi}_{\sigma},\hat{H} \end{bmatrix},
\end{equation}
and applying the mean-field decoupling of the cubic terms in $\hat{\Psi}$ and $\hat{\Psi}^\dagger$~\cite{griffin1996conserving}, we obtain two coupled equations as in Refs.~\cite{chen2017quantum,chen2018quantum}.

The first equation is the modified Gross-Pitaevskii (GP) equation for the condensate,
\begin{equation}  \label{eq:gp}
\left[\mathcal{H}_{s}(\hat{{\bf p}})+g_{_{\uparrow\downarrow}}n_\mathrm{sf}\sigma_x+\mathrm{diag}(\mathcal{L}_\uparrow,\mathcal{L}_\downarrow)\right] \psi= \mu \psi
\end{equation}
with the spinor $\psi\equiv\left(\psi_{\uparrow},\psi_{\downarrow}\right)^T$, chemical potential $\mu$, the spin-flip density term $n_\mathrm{sf}$ and the diagonal element $\mathcal{L}_{\sigma}\equiv g(n_{c\sigma}+2n_{t\sigma})+g_{_{\uparrow\downarrow}}n_{\bar{\sigma}}$ (here $\bar{\sigma}\neq\sigma$). The second equation is the coupled Bogoliubov equation for quasiparticles,
\begin{subequations}  \label{eq:bogoliubov}
\begin{eqnarray}
\left[\mathcal{H}_{s}(\hat{\bf p})-\mu+\mathcal{A}_\uparrow\right]U_j +\mathcal{B}V_j&=&\omega_jU_j, \\
-\mathcal{B}U^*_j -\left[\mathcal{H}_{s}(\hat{\bf p})-\mu+\mathcal{A}^*_\downarrow\right]V^*_j&=&\omega_jV^*_j,
\end{eqnarray}
\end{subequations}
where $U_j\equiv\left(u_{j\uparrow}, u_{j\downarrow}\right)^T$, $V_j\equiv\left(v_{j\uparrow}, v_{j\downarrow}\right)^T$, and
\begin{subequations}  
\begin{equation}
\mathcal{A}_\sigma\equiv 
\begin{pmatrix}
2gn_{\uparrow}+g_{_{\uparrow\downarrow}}n_{\downarrow} &g_{_{\uparrow\downarrow}}\left(\psi_{\sigma}\psi_{\bar{\sigma}}^*+n_\mathrm{sf}\right)\\ g_{_{\uparrow\downarrow}}\left(\psi_{\bar{\sigma}}\psi_{\sigma}^*+n_\mathrm{sf}\right) &2gn_{\downarrow}+g_{_{\uparrow\downarrow}}n_{\uparrow}
\end{pmatrix},
\end{equation}
\begin{equation} 
\mathcal{B}\equiv 
\begin{pmatrix}
g\phi^2_{\uparrow} &g_{_{\uparrow\downarrow}}\psi_{\uparrow}\psi_{\downarrow}\\ g_{_{\uparrow\downarrow}}\psi_{\uparrow}\psi_{\downarrow} &g\psi^2_{\downarrow}
\end{pmatrix}.
\end{equation}
\end{subequations}

In these equations, $n_{c\sigma}$ is the condensate density of spin components $\sigma=\uparrow,\downarrow$ and we have applied the Popov approximation to ensure a gapless theory~\cite{griffin1996conserving,chen2015collective}, i.e., omitting the anomalous densities $\langle\hat{\eta}^{\dagger}\hat{\eta}^{\dagger}\rangle$ and $\langle\hat{\eta}\hat{\eta}\rangle$. To take quantum and thermal fluctuations into account, we have introduced a non-condensate density $n_{t\sigma}\equiv\langle\hat{\eta}_{\sigma}^{\dagger}\hat{\eta}_{\sigma}\rangle=(1/V)\sum_{{\bf q\lambda}}[(|u_{{\bf q}\sigma}^{(\lambda)}|^{2}+|v_{{\bf q}\sigma}^{(\lambda)}|^{2})/(e^{\beta\omega_{{\bf q}\lambda}}-1)+|v_{{\bf q}\sigma}^{(\lambda)}|^{2}]$ and a spin-flip density term $n_\mathrm{sf} \equiv\langle\hat{\eta}_{\uparrow}^{\dagger}\hat{\eta}_{\downarrow}\rangle=(1/V)\sum_{{\bf q\lambda}}[(u_{{\bf q}\uparrow}^{(\lambda)}u_{{\bf q}\downarrow}^{(\lambda)}+v_{{\bf q}\uparrow}^{(\lambda)}v_{{\bf q}\downarrow}^{(\lambda)})/(e^{\beta\omega_{{\bf q}\lambda}}-1)+v_{{\bf q}\uparrow}^{(\lambda)}v_{{\bf q}\downarrow}^{(\lambda)}]$ at a temperature $\beta\equiv1/(k_{B}T)$. Thus, the total density of spin $\sigma$ is given as $n_{\sigma}=n_{c\sigma}+n_{t\sigma}$. 

At zero temperature, the thermal part of $n_{t\sigma}$ and the spin-flip term $n_\mathrm{sf}$ vanish and we can safely neglect them. Nevertheless, the condensate is still depleted by a small fraction of the total density, even at zero temperature, due to quantum fluctuations. This is the so-called quantum depletion, $n_{\mathrm{qd}}=\sum_{{\bf q}\lambda\sigma}|v_{{\bf q}\sigma}^{(\lambda)}|^{2}/V$, involving typically about $1\%$ of the total density in the weakly-interacting regime, i.e., $n_{c\uparrow}+n_{c\downarrow}\simeq n$. In our previous work, we have shown that the SOC effect does not affect remarkably the quantum depletion fraction in the condensate density (see Fig. 4 of Ref.~\cite{chen2018quantum}).

\subsection{The {\it Ansatz} and ground-state phases}

We aim to solve the modified GP equation and the coupled Bogoliubov equations at finite temperature and then derive the current-current response functions. Before that, for self-containedness let us briefly review the phase diagram at zero temperature, where the model Hamiltonian of a Raman-type SOC Bose gas in Sec.~\ref{sec:model} can be solved straightforwardly using a variational formalism.

To obtain the ground-state phases, a variational \emph{first-order} {\it stripe Ansatz} is usually taken as~\cite{li2012quantum,martone2014approach,yu2014equation,chen2018quantum}
\begin{equation} \label{eq:1st_stripe}
\psi(\mathbf{r})=\sqrt{n}\left[C_1\begin{pmatrix}
\sin\theta\\
-\cos\theta
\end{pmatrix} e^{-iP_{x}x}+
C_2\begin{pmatrix}
\cos\theta\\
-\sin\theta
\end{pmatrix} e^{iP_{x}x}\right],
\end{equation}
with the uniform average density $n=N/V$, and the variational angle $\theta$ in the range $[0,\pi/4]$ weighing the spin components of the condensate.  By substituting this trial wave function, Eq.~\eqref{eq:1st_stripe}, into the model Hamiltonian, the total energy of the system can be written in terms of $\psi({\bf r})$ as
\begin{equation} \label{eq:mean-field-energy}
\begin{aligned}
E= & \int d^{3}{\bf r}\left[\begin{pmatrix}
\psi_{\uparrow}^*({\bf r}),\psi_{\downarrow}^*({\bf r})
\end{pmatrix}\mathcal{H}_{\mathrm{s}}(\hat{{\bf p}})
\begin{pmatrix}
\psi_{\uparrow}({\bf r})\\
\psi_{\downarrow}({\bf r})
\end{pmatrix}\right.\\
 & \left.+\frac{1}{2}g\left(|\psi_{\uparrow}({\bf r})|^{4}+|\psi_{\downarrow}({\bf r})|^{4}\right)+g_{\uparrow\downarrow}|\psi_{\uparrow}({\bf r})|^{2}|\psi_{\downarrow}({\bf r})|^{2}\right].
\end{aligned}
\end{equation}
After minimizing the ground-state energy with respect to the variational parameters $C_{1,2}$, $\theta$, and $P_x$, one can in general find three exotic phases, i.e., the stripe (ST), plane-wave (PW), and zero-momentum (ZM) phases, in the appropriate interaction regimes with $g>g_{\uparrow\downarrow}$. This condition is necessary for the existence of the exotic stripe phase~\cite{li2012quantum}. In addition, by defining two interaction parameters $G_{1}=(g+g_{\uparrow\downarrow})n/4$ and $G_{2}=(g-g_{\uparrow\downarrow})n/4$, the critical Rabi frequency $\Omega$ of three phases can be determined respectively by~\cite{li2012quantum}
\begin{subequations}
\begin{equation} \label{eq:omega1}
\Omega_{c1}=2\left[(2E_{\mathrm{r}}+G_{1})(2E_{\mathrm{r}}-2G_{2})\frac{2G_{2}}{G_{1}+2G_{2}}\right]^{1/2}
\end{equation}
for the ST-PW phase transition, and 
\begin{equation} \label{eq:omega2}
\Omega_{c2}=4E_{\mathrm{r}}-4G_{2}
\end{equation}
\end{subequations}
for the PW-ZM phase transition. As illustrated in our previous work~\cite{chen2017quantum}, these critical positions can be notably altered by the quantum and thermal fluctuations which favor the existence of PW phase. It is worth mentioning that, the {\it Ansatz} of Eq.~\eqref{eq:1st_stripe} is a superposition of two plane waves with momenta $\pm P_{x}$ without including any higher-order terms. Higher-order harmonics could be non-negligible when the interaction energies $G_{1,2}$ become relatively large~\cite{chen2018quantum}. For example, these high-order contributions will shift the ST-PW phase transition point $\Omega_{c1}$ to a larger Rabi frequency, since the higher-order stripe {\it Ansatz} can host lower energy solutions than that of the PW phase.

The consideration of the stripe {\it Ansatz} at finite temperature is much more involved and is beyond the scope of this work. Here, we focus on the plane-wave and zero-momentum phases, which are characterized by a {\it single-plane-wave Ansatz} with a condensate density $n_{c}$ at a momentum $P_{x}$, and are obtained by taking $C_1=0$ or $C_2=0$ in Eq.~\eqref{eq:1st_stripe}~\cite{martone2012anisotropic,zheng2013properties,zhang2016superfluid,chen2017quantum,chen2018quantum}:
\begin{equation} \label{eq:plane-wave}
\psi({\bf r})=
\begin{pmatrix}
\psi_{\uparrow} \\ \psi_{\downarrow}
\end{pmatrix} e^{iP_{x}x}=\sqrt{n_{c}}\left(\begin{array}{c}
\cos\theta \\ -\sin\theta
\end{array}\right)e^{iP_{x}x}.
\end{equation}
With this trial {\it Ansatz}, the minimization of the mean-field energy gives rise to two solutions with respective to Rabi frequency~\cite{li2012quantum,martone2012anisotropic,zheng2013properties,chen2017quantum}. At $\Omega<\Omega_{c2}$, the magnetic plane-wave phase appears with a nonzero momentum $P_{x}=\pm k_{\mathrm{r}}\sqrt{1-\Omega^{2}/\Omega_{c2}^{2}}$ occupied by the condensates,  a nonzero magnetization $\langle\sigma_z\rangle\neq0$, and a typical roton-maxon structure in the lowest-lying excitation spectrum. In contrast, the nonmagnetic zero-momentum phase occurs for $\Omega\geq\Omega_{c2}$ where the condensate momentum and the magnetization are both zero, i.e., $P_{x}=0$ and $\langle\sigma_z\rangle=0$, and there exist only the conventional linear phonon modes in the long-wavelength limit in the lowest-lying excitation spectrum.

\subsection{The PW and ZM phases at finite temperature}

At a certain finite temperature $T$ and a Rabi frequency $\Omega$, we can also take this plane-wave {\it Ansatz} in terms of variational parameters ($\theta,P_{x}$), and solve self-consistently the GP and Bogoliubov equations in Eqs.~\eqref{eq:gp} and \eqref{eq:bogoliubov}. Thus, the free energy of the system, i.e., $\mathcal{F}(\theta,P_{x})$ in Ref.~\cite{chen2017quantum}, can be calculated straightforwardly. The condensate wave function $\psi({\bf r})$, the Bogoliubov quasiparticle wave functions $u_{j}(v_{j})$ and excitation spectrum $\omega_{j}$ can be then determined by minimizing the free energy $\mathcal{F}$ with respect to two variational parameters, namely, $(\partial\mathcal{F}/\partial\theta)_{N}=0$ and $(\partial\mathcal{F}/\partial P_{x})_{N}=0$.

The recent experiments have successfully utilized ultracold $^{87}$Rb atoms to realize this Raman-type SOC~\cite{lin2011spin,ji2014experimental,ji2015softening}. For instance in Ref.~\cite{ji2015softening}, the typical interaction energy is $gn=0.38E_{\mathrm{r}}$ with a peak density $n=0.46k_{\mathrm{r}}^{3}$ in harmonic traps, and the ratio $g_{_{\uparrow\downarrow}}/g=100.99/101.20$ between the inter- and intra-species interactions. Accordingly, the two critical Rabi frequencies characterizing the ST-PW and PW-ZM phase transitions at zero temperature, can be determined respectively from Eqs.~\eqref{eq:omega1} and~\eqref{eq:omega2}, i.e., $\Omega_{c1}=0.2E_{\mathrm{r}}$ and $\Omega_{c2}=4.0E_{\mathrm{r}}$. Previous works have shown that the regime of the stripe phase (i.e., $\Omega_{c1}$) can be actually tuned by the difference between intra- and inter-species interactions~\cite{li2012quantum,martone2014approach}. As we are concentrating on the plane-wave and zero-momentum phases in this work, we will set $g_{_{\uparrow\uparrow}}=g_{_{\downarrow\downarrow}}=g\geq g_{_{\uparrow\downarrow}}$ in the following calculations. Also, we will consider a relatively small difference between inter- and intra-species interaction strengths, i.e., zero or relatively small $G_{2}$, in order to enlarge the window for the plane-wave and zero-momentum phases in the phase diagram.

\section{Results and Discussions\label{sec:results}}

We are now ready to perform our derivations of the current-current response functions and extract the superfluid density at finite temperature. To address the effect of the one-dimensional Raman-type SOC, we will focus only on the SOC direction (i.e., in the $x$-axis) in the following. In the perpendicular plane, the SOC does not affect the elementary excitations in the long-wavelength limit and the superfluid density fraction remains unity at zero temperature (see Refs.~\cite{zhang2016superfluid,chen2018quantum}). 

\subsection{Current-current response functions\label{sec:JJ-response-function}}

In this subsection, we study the current-current response functions that are closely associated with the superfluidity of the SOC Bose gas. Technically, the superfluid density can be calculated by the difference between the longitudinal and transverse current-current response functions in the long-wavelength and low-frequency limits~\cite{griffin1993excitations,pitaevskii2016bose,nozieres2018theory}.

We start by finding the expression for the current operator in the presence of the Raman-type SOC. From the equation of motion in Eq.~\eqref{eq:eq-of-motion} for the Bose field operators, we obtain the continuity equation~\cite{stringari2017diffused,tang2018quantum} \begin{eqnarray} 
\partial_tn+\nabla\cdot (J^0-\frac{2k_\mathrm{r}}{m}s_z\hat{\bf e}_x)&=&0. \end{eqnarray}
Here $J^0=J^0_{\uparrow\uparrow}+J^0_{\downarrow\downarrow}$ is the conventional total-density current with the current operators $\hat{J}^0_{\mu\nu}\equiv\frac{1}{2mi}(\hat{\Psi}_{\mu}^*\nabla\hat{\Psi}_{\nu}-\nabla\hat{\Psi}_{\mu}^*\hat{\Psi}_{\nu})$ in the absence of SOC, and $\hat{s}_{z}\equiv\frac{1}{2}\sum_{\mu,\nu}\hat{\Psi}_{\mu}^*\hat{\sigma}_z^{\mu\nu}\hat{\Psi}_{\nu}$ is the spin density. It is clear that due to SOC the total-density current is coupled to the spin density. Therefore, it is necessary to consider the modified current operators for each spin component $\sigma=(\uparrow,\downarrow)$, 
\begin{subequations} \label{eq:J-density-operator} \begin{eqnarray} 
\hat{j}_{\uparrow\uparrow}({\bf r})&=&\frac{1}{2mi}(\hat{\Psi}_{\uparrow}^*\nabla_x\hat{\Psi}_{\uparrow}-\nabla_x\hat{\Psi}_{\uparrow}^*\hat{\Psi}_{\uparrow})-\frac{k_\mathrm{r}}{m}\hat{\Psi}_{\uparrow}^*\hat{\Psi}_{\uparrow}, \\ 
\hat{j}_{\downarrow\downarrow}({\bf r})&=&\frac{1}{2mi}(\hat{\Psi}_{\downarrow}^*\nabla_x\hat{\Psi}_{\downarrow}-\nabla_x\hat{\Psi}_{\downarrow}^*\hat{\Psi}_{\downarrow})+\frac{k_\mathrm{r}}{m}\hat{\Psi}_{\downarrow}^*\hat{\Psi}_{\downarrow}, 
\end{eqnarray} 
\end{subequations}
such that the total-density current $\hat{j}_d({\bf r}) \equiv \hat{j}_{\uparrow\uparrow}+\hat{j}_{\downarrow\downarrow}$ reads
\begin{eqnarray} \label{eq:Jd} 
\hat{j}_d({\bf r}) &=& \frac{1}{2mi}\sum_{\mu}(\hat{\Psi}_{\mu}^*\nabla_x\hat{\Psi}_{\mu}-\nabla_x\hat{\Psi}_{\mu}^*\hat{\Psi}_{\mu})-\frac{2k_\mathrm{r}}{m}\hat{s}_z. 
\end{eqnarray}

We expand the spin-component current operators using the well-defined Bose field operator in Eq.~\eqref{eq:newBosefield}, and the operators are divided into three parts
\begin{eqnarray}
\hat{j}_{\sigma\sigma}({\bf r},t)&\equiv&j_{0\sigma}({\bf r})+\delta\hat{j}_{\sigma}({\bf r},t)+\tilde{j}_{\sigma}({\bf r},t),
\end{eqnarray}
where $j_{0\sigma}$, $\delta\hat{j}_{\sigma}$, and $\tilde{j}_{\sigma}$ represent the zeroth-, first-, and second-order terms of the fluctuation operator $\hat{\eta}$, respectively, being
\begin{subequations}
\begin{eqnarray}
j_{0\sigma}({\bf r}) &=& \frac{1}{2mi}(\psi_{\sigma}^*\nabla_x\psi_{\sigma}-\nabla_x\psi_{\sigma}^*\psi_{\sigma})\mp\frac{k_\mathrm{r}}{m}\psi_{\sigma}^*\psi_{\sigma},
\\
\delta\hat{j}_{\sigma}({\bf r},t) &=& \frac{1}{2mi}(\psi_{\sigma}^*\nabla_x\hat{\eta}_{\sigma}+\nabla_x\psi_{\sigma}\hat{\eta}_{\sigma}^*-\nabla_x\psi_{\sigma}^*\hat{\eta}_{\sigma}-\psi_{\sigma}\nabla_x\hat{\eta}_{\sigma}^*) \nonumber \\
&& \mp\frac{k_\mathrm{r}}{m}(\psi_{\sigma}^*\hat{\eta}_{\sigma}+\hat{\eta}_{\sigma}^*\psi_{\sigma}), \label{eq:delta-J}
\\
\tilde{j}_{\sigma}({\bf r},t) &=& \frac{1}{2mi}(\hat{\eta}_{\sigma}^*\nabla_x\hat{\eta}_{\sigma}-\nabla_x\hat{\eta}_{\sigma}^*\hat{\eta}_{\sigma})\mp\frac{k_\mathrm{r}}{m}\hat{\eta}_{\sigma}^*\hat{\eta}_{\sigma}.
\end{eqnarray}
\end{subequations}
Here the sign '$\mp$' is $-$ ($+$) for spin component $\sigma=\uparrow$ ($\downarrow$).  

Within the Bogoliubov approximation, we are interested only in the current fluctuation (i.e., $\delta \hat{j}$) or the linear terms of the fluctuation operator $\hat{\eta}$~\cite{griffin1993excitations}. Therefore, within the imaginary-time (i.e., $\tau=it$) Green’s function method~\cite{fetter2003quantum,mahan2013many,liu2004collective}, the time-ordered correlation function $\langle T_{\tau}\delta\hat{j}_{\sigma}({\bf r},\tau)\delta\hat{j}_{\sigma^\prime}({\bf r^\prime})\rangle$ for the current fluctuation density operator can be derived explicitly in terms of the condensate wave function $\psi$ and the Bogoliubov quasiparticle wave functions $u$($v$). The details can be seen in Appendix~\ref{app:j-j-response}. We summarize the main results as follows.

We take a {\it  plane-wave Ansatz} in Eq.~\eqref{eq:plane-wave} for the classic condensate wave function and the time-ordered correlation function of the current fluctuation density can be consequently rewritten as 
\begin{equation} \label{eq:jj-correlation-new}
\langle T_{\tau}\delta\hat{j}_{\sigma}({\bf r},\tau)\delta\hat{j}_{\sigma^\prime}({\bf r^\prime})\rangle  \equiv \mathcal{A}_1+\mathcal{A}_2+\mathcal{A}_3+\mathcal{A}_4, 
\end{equation}
in terms of four correlation functions $\mathcal{A}_{m}$ $(m=1,2,3,4)$ for the fluctuation operator $\hat{\eta}$, which are given explicitly in Appendix~\ref{app:j-j-response}. Thus, the response functions of the elements in $\mathcal{A}_m$ can be calculated by \begin{equation} 
\chi^{\mu\nu}_{m}({\bf r},{\bf r^\prime};i\omega_n) = \int_0^\beta d\tau e^{i\omega_n\tau}\frac{\theta(\tau)}{V} \mathcal{A}_m^{\mu\nu}({\bf r},{\bf r^\prime};\tau) \nonumber 
\end{equation}
with a step function $\theta(\tau)$, the system volume $V$ and the Matsubara frequencies $i\omega_n\equiv2n\pi i/\beta$ for bosons. The current response function is then calculated collectively by using Eq.~\eqref{eq:delta-J} as
\begin{eqnarray}  \label{eq:chi_JJ_rr} &&\chi_{JJ}({\bf r},{\bf r^\prime};i\omega_n)\equiv\chi^{\uparrow\uparrow}+\chi^{\uparrow\downarrow}+\chi^{\downarrow\uparrow}+\chi^{\downarrow\downarrow} \nonumber \\ 
&=& \int_0^\beta d\tau e^{i\omega_n\tau} \theta(\tau)\langle T_{\tau}\delta\hat{j}({\bf r},\tau)\delta\hat{j}({\bf r^\prime})\rangle. \end{eqnarray}
By employing the analytic continuation $(i\omega_n\rightarrow\omega+i\eta)$ with $\eta=0^+$, setting $\omega=0$ and  then taking the Fourier transform to momentum space, the ${\bf q}$-component static response function is obtained in terms of the elements $\chi^{\sigma\sigma^\prime}({\bf q};\omega=0)$, i.e.,
\begin{subequations} \label{eq:ssprime-response}
\begin{eqnarray}
\chi^{\uparrow\uparrow}
&=&\frac{\psi_{\uparrow}^2}{m^2V}\sum_{j} \frac{C_1u^2_{\uparrow}+C_2v^2_{\uparrow}+C_3u_{\uparrow}v_{\uparrow}}{\omega_j}, 
\\
\chi^{\downarrow\downarrow}
&=&\frac{\psi_{\downarrow}^2}{m^2V}\sum_{j}\frac{C_4u^2_{\downarrow}+C_5v^2_{\downarrow}+C_6u_{\downarrow}v_{\downarrow}}{\omega_j},
\\
\chi^{\uparrow\downarrow}&=&\chi^{\downarrow\uparrow} = \frac{\psi_{\uparrow}\psi_{\downarrow}}{m^2V}\cdot\sum_{j} \nonumber\\
&& \frac{C_7u_{\uparrow}u_{\downarrow}+C_8v_{\uparrow}v_{\downarrow}+C_9u_{\uparrow}v_{\downarrow}+C_{10}v_{\uparrow}u_{\downarrow}}{\omega_j},
\end{eqnarray}
\end{subequations}
where the index $j$ runs over all the possible single-particle states. Without confusion, we have made the index $j$ implicit in the quasiparticle wave-functions $u$ and $v$, i.e., $u_{\sigma}\equiv u_{j,\sigma}$ and $v_{\sigma}\equiv v_{j,\sigma}$ are calculated at each momentum ${\bf q}$ for the energy level $\omega_j$ (see the notations in Sec.~\ref{sec:bogoliubov}). The coefficients $C_n$ are given by
\begin{subequations}
\begin{eqnarray}
C_{1/2} &=& \frac{1}{2}(2P_x\pm q_x-2k_\mathrm{r})^2, 
\\
C_{3/6} &=& (2P_x-q_x\mp 2k_\mathrm{r})(2P_x+q_x\mp 2k_\mathrm{r}),  
\\
C_{4/5} &=& \frac{1}{2}(2P_x\pm q_x+2k_\mathrm{r})^2,  
\\
C_{7/8} &=& \frac{1}{2}(2P_x\pm q_x)^2-2k_\mathrm{r}^2, 
\\
C_{9/10} &=& 2P_x^2-\frac{1}{2}(q_x\mp 2k_\mathrm{r})^2.
\end{eqnarray}
\end{subequations}
The explicit derivation can be seen  in Appendix~\ref{app:j-j-response}. Particularly in the zero-momentum phase with the condensates located at $P_x=0$, the coefficients reduce to
\begin{eqnarray*}
C^{(\mathrm{ZM})}_{1/2} &=& C^{(\mathrm{ZM})}_{5/4} = -C^{(\mathrm{ZM})}_{9/10} = \frac{1}{2}(q_x\mp 2k_\mathrm{r})^2, 
\\
C^{(\mathrm{ZM})}_3 &=& C^{(\mathrm{ZM})}_6 = -2C^{(\mathrm{ZM})}_7 = -2C^{(\mathrm{ZM})}_8 = 4k_\mathrm{r}^2-q_x^2.
\end{eqnarray*}
Finally, the ${\bf q}$-component static current-current response function is then obtained by the summation
\begin{equation}
\chi_{JJ}({\bf q}; \omega=0)=\sum_{\sigma,\sigma^\prime} \chi^{\sigma\sigma^\prime}({\bf q};\omega=0).
\end{equation}
\begin{figure}[t]
\includegraphics[width=0.48\textwidth]{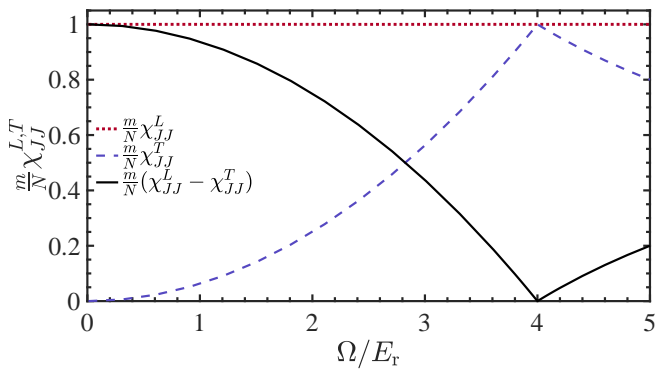}
\caption{The longitudinal and transverse (i.e., dotted red and dashed blue lines, respectively) static current-current response functions $\frac{m}{N}\chi^{L,T}_{JJ}({\bf q};0)$ in the ${\bf q}\to0$ limit, as a function of Rabi frequency $\Omega$. Their difference (i.e., solid black line) reveals the superfluid density fraction. Here we take the parameters as in the experiment~\cite{ji2015softening}, with a total density $n=0.46k_\mathrm{r}^3$, $gn=0.38E_\mathrm{r}$ and $g_{_{\uparrow\downarrow}}/g=100.99/101.20$, i.e., $G_1=0.19E_\mathrm{r}$ and $G_2\approx0$.}
\label{fig1}
\end{figure}

We consider only the SOC direction (i.e., $x$ axis) and the corresponding total and normal densities can be then calculated via the longitudinal and transverse components of the total-current response function $\chi_{JJ}({\bf q};0)$ in the long wavelength limit~\cite{baym1968microscopic,pitaevskii2016bose}. Therefore, the {\it superfluid mass density} along the SOC direction is given by
\begin{equation} \label{eq:ns-JJ-response}
\rho_s = \rho-\rho_n = \rho\frac{m}{N}(\lim_{q_x\to0}\lim_{q_{\perp}\to0}\chi_{JJ}-\lim_{q_{\perp}\to0}\lim_{q_x\to0}\chi_{JJ}).
\end{equation}

An example of the current-current response function of a zero-temperature SOC $^{87}$Rb Bose gas in the static and low-wavelength limits is illustrated in Fig.~\ref{fig1}, with a total density $n=0.46k_\mathrm{r}^3$, the interaction energy $gn=0.38E_\mathrm{r}$ and the ratio of inter-, intra-species interaction strength $g_{_{\uparrow\downarrow}}/g=100.99/101.20$ taken from a recent experiment~\cite{ji2015softening}. The longitudinal static current-current response function or the total density (i.e., dotted red line) remains unity while the transverse component or the normal density fraction (i.e., dashed blue line) changes non-monotonically. Their difference (i.e., solid black line) reveals the superfluid density fraction, which is significantly affected by the SOC effect. The zero-temperature superfluid fraction in Fig.~\ref{fig1}, explicitly calculated using the current-current response functions, agrees with the earlier result obtained via a sum-rule approach~\cite{zhang2016superfluid}.

\subsection{Generalized Josephson relation \label{sec:Josephson}}

Superfluid density is intimately related to Bose-Einstein condensate density. Nonetheless, they are not exactly equivalent even at zero temperature, for instance in liquid helium with strong interactions~\cite{penrose1956bose}, or in weakly-interacting Bose gases with SOC~\cite{zhang2016superfluid,chen2018quantum}. In 1965, B.D. Josephson and other researchers have derived the so-called Josephson relation or Josephson sum rule for a conventional weakly-interacting Bose gas, which associates the superfluid mass density $\rho_s$ with the condensate density $n_0$ at zero or finite temperatures as~\cite{hohenberg1965microscopic,josephson1966relation,baym1968microscopic,holzmann2007condensate}
\begin{equation} \label{eq:josephson_relation}
    \rho_s = -\lim_{{\bf q}\to0}\frac{n_0m^2}{{\bf q}^2G_{11}({\bf q},0)},
\end{equation}
with the atomic mass $m$ and the single-particle Green's function at momentum ${\bf q}$ and zero frequency. It is straightforward to show that, for the conventional weakly-interacting Bose gas at zero temperature, within the Bogoliubov approximation the superfluid mass density is connected to the condensate density via the Josephson relation~\cite{griffin1993excitations}. Here, we are interested in examining the Josephson relation within the same Bogoliubov approximation at exactly \emph{zero} temperature. The situation at nonzero temperature is more involved, as accurate calculations of the superfluid density and condensate fraction may require different levels of approximation. Therefore, we defer to a future study for the nonzero-temperature case.

For the examination at zero temperature, let us first discuss the Josephson relation in the presence of SOC. In 2018, Yi-Cai Zhang generalizes the Josephson relation to multi-component Bose gases using the Green's function within the linear response theory~\cite{zhang2018generalized}. In this work, we will follow his procedure and consider a two-component SOC Bose gas. We focus only on the SOC direction, i.e., the $x$-axis. Thus, in terms of the order parameter $\langle \hat{\Psi}\rangle$ and Green's function matrix in the long-wavelength limit, the expression for the superfluid density in the SOC direction is given by~\cite{zhang2018generalized}
\begin{equation}  \label{eq:ns-Josephson}
\rho_{\mathrm{s}}^\mathrm{(SOC)} = -\lim_{{\bf q}\rightarrow0}\frac{1}{q^2} 
\begin{pmatrix}
\langle \hat{\Psi}\rangle, \langle \hat{\Psi}^*\rangle
\end{pmatrix} \sigma^\mathbb{Z} \mathcal{G}^{-1}\sigma^\mathbb{Z}
\begin{pmatrix}
\langle \hat{\Psi}\rangle  \\
   \langle \hat{\Psi}^*\rangle
\end{pmatrix}.
\end{equation}
Here we have introduced a $4\times4$ matrix $\sigma^\mathbb{Z}\equiv \left(\begin{array}{cc} \mathbb{I} & 0 \\ 0 & -\mathbb{I} \end{array}\right)$ and a $2\times2$ identity matrix $\mathbb{I}$. The Green's matrix is defined as
\begin{eqnarray} 
\mathcal{G} &\equiv& 
\begin{pmatrix}
G({\bf q},0)  & F^T({\bf q},0) \\
   (F^T({\bf q},0))^* & G^*({\bf q},0)
\end{pmatrix},
\end{eqnarray} 
and
\begin{subequations} \label{eq:G-and-F}
\begin{eqnarray}
G({\bf q},0) &\equiv& 
\begin{pmatrix}
G_{\uparrow\uparrow}({\bf q},0)  & G_{\uparrow\downarrow}({\bf q},0) \\
   G_{\downarrow\uparrow}({\bf q},0) & G_{\downarrow\downarrow}({\bf q},0)
\end{pmatrix}, \\
F({\bf q},0) &\equiv& 
\begin{pmatrix}
F_{\uparrow\uparrow}({\bf q},0)  & F_{\uparrow\downarrow}({\bf q},0) \\
   F_{\downarrow\uparrow}({\bf q},0) & F_{\downarrow\downarrow}({\bf q},0)
\end{pmatrix}.
\end{eqnarray}
\end{subequations}
Here $G_{\sigma\sigma'}$ and $F_{\sigma\sigma'}$ are normal and anomalous Green’s function matrix elements. It is worth noting that the above derivations are assumed for homogeneous systems, which possess the translational symmetry. 

Let us now derive the expression for Green functions. In the presence of the Raman-type SOC, one can start with a {\it plane-wave Ansatz} for the condensate wave function in the plane-wave and zero-momentum phases~\cite{martone2012anisotropic,zheng2013properties,chen2017quantum,chen2018quantum}, i.e.,  $\psi({\bf r})= \begin{pmatrix} \psi_{\uparrow}  \\ \psi_{\downarrow} \end{pmatrix} e^{iP_xx}$ as in Eq.~\eqref{eq:plane-wave}, satisfying the Gross-Pitaevskii equation with the condensation momentum $P_x\geq0$ and the average density $n=N/V$. To the next order, i.e., within the Bogoliubov approximation, the Hamiltonian for the fluctuation operators is written as~\cite{zheng2013properties}
\begin{eqnarray}
    \hat{H}_\mathrm{bogo} &=& \frac{1}{2}\sum_{{\bf q}}\hat{\Phi}^\dagger_{{\bf q}}H_{\mathrm{B}}({\bf q})\hat{\Phi}_{{\bf q}}+\epsilon_\mathrm{shift},
\end{eqnarray}
with $\epsilon_\mathrm{shift}=- \frac{1}{2}\left[\xi_{_{P_x-{\bf q},\uparrow}}+\xi_{_{P_x-{\bf q},\uparrow}}-2\mu+2gn+g_{_{\uparrow\downarrow}}n\right]$ being an energy shift arising from the bosonic commutation relations and $\xi_{\bf k}\equiv {\bf k}^2/(2m)$. Here $H_{\mathrm{B}}$ is the Bogoliubov Hamiltonian for the $4$-component Nambu spinor $\hat{\Phi}_{{\bf q}}\equiv[\hat{\phi}_{P_x+{\bf q},\uparrow},\hat{\phi}_{P_x+{\bf q},\downarrow},\hat{\phi}^\dagger_{P_x-{\bf q},\uparrow},\hat{\phi}^\dagger_{P_x-{\bf q},\downarrow}]^T$ as
\begin{equation}
H_{\mathrm{B}}({\bf q}) \equiv
\begin{bmatrix}
K_0(P_x+{\bf q})+\Sigma_\mathrm{N} & \Sigma_\mathrm{A}  \\
         \Sigma_\mathrm{A} & K_0(P_x-{\bf q})+\Sigma_\mathrm{N}
\end{bmatrix},
\end{equation}
where we have defined three matrices 
\begin{subequations}
\begin{eqnarray}
K_0({\bf q})   &=& 
\begin{bmatrix}
\frac{(q_x-k_\mathrm{r})^2+q_{\perp}^2}{2m}-\mu & \frac{\Omega}{2}\\   
    \frac{\Omega}{2} & \frac{(q_x+k_\mathrm{r})^2+q_{\perp}^2}{2m}-\mu
\end{bmatrix}, 
\\
\Sigma_\mathrm{N} &=& 
\begin{bmatrix}
2gn_{\uparrow}+g_{_{\uparrow\downarrow}}n_{\downarrow} & g_{_{\uparrow\downarrow}}\psi_{\uparrow}\psi_{\downarrow} \\   
    g_{_{\uparrow\downarrow}}\psi_{\uparrow}\psi_{\downarrow} & 2gn_{\downarrow}+g_{_{\uparrow\downarrow}}n_{\uparrow} 
\end{bmatrix}, 
\\
\Sigma_\mathrm{A} &=& 
\begin{bmatrix}
gn_{\uparrow} & g_{_{\uparrow\downarrow}}\psi_{\uparrow}\psi_{\downarrow} \\  
    g_{_{\uparrow\downarrow}}\psi_{\uparrow}\psi_{\downarrow} & gn_{\downarrow} 
\end{bmatrix},
\end{eqnarray}
\end{subequations}
with the spin density $n_{\sigma}\equiv|\psi_{\sigma}|^2$ and $q_{\perp}^2\equiv q_{y}^2+q_{z}^2$.

The gapless condition $\mathrm{Det}[K_0(P_x)+\Sigma_\mathrm{N}-\Sigma_\mathrm{A}]=0$ at ${\bf q}=0$ is ensured by the GP equation, with which the chemical potential can be solved as $\mu=(P_x^2+k_\mathrm{r}^2)/2m+(g+g_{_{\uparrow\downarrow}})n/2-\sqrt{\left(-P_xk_\mathrm{r}/m+(g-g_{_{\uparrow\downarrow}})(n_{\uparrow}-n_{\downarrow})/2\right)^2+\Omega^2/4}$. Hence, the diagonal elements of $K_0(P_x+{\bf q})$ can be rewritten in terms of the explicit chemical potential and we can write straightforwardly the inverse single-particle Green's function matrix with SOC as
\begin{eqnarray}
&& G^{-1}_{\mathrm{B}}({\bf q},i\omega_n) = \nonumber \\
&&\begin{bmatrix}
i\omega_n\mathbb{I}-K_0(P_x+{\bf q})-\Sigma_\mathrm{N} & -\Sigma_\mathrm{A}  \\
-\Sigma_\mathrm{A} & -i\omega_n\mathbb{I}-K_0(P_x-{\bf q})-\Sigma_\mathrm{N}
\end{bmatrix}_{4\times4}.
\end{eqnarray}

In particular, in the zero-momentum phase with $P_x=0$ and $\psi_{\uparrow}=-\psi_{\downarrow}=\sqrt{n/2}$, the chemical potential reduces to $\mu^\mathrm{ZM}=k_\mathrm{r}^2/2m+(g+g_{_{\uparrow\downarrow}})n/2-\Omega/2$, and the relevant matrices become $K^\mathrm{ZM}_0({\pm}{\bf q})  =
\begin{bmatrix}
\frac{{\pm}2k_\mathrm{r}q_x+{\bf q}^2}{2m}-2G_1+\frac{\Omega}{2} & \frac{\Omega}{2}\\   
   \frac{\Omega}{2} & \frac{{\mp}2k_\mathrm{r}q_x+{\bf q}^2}{2m}-2G_1+\frac{\Omega}{2}
\end{bmatrix}$
and
\begin{subequations}
\begin{eqnarray}
\Sigma^\mathrm{ZM}_\mathrm{N} &=& \frac{1}{2}
\begin{bmatrix}
2gn+g_{_{\uparrow\downarrow}}n & -g_{_{\uparrow\downarrow}}n \\   
   -g_{_{\uparrow\downarrow}}n & 2gn+g_{_{\uparrow\downarrow}}n
\end{bmatrix},
\\
\Sigma^\mathrm{ZM}_\mathrm{A} &=& \frac{1}{2}
\begin{bmatrix}
gn & -g_{_{\uparrow\downarrow}}n \\   
   -g_{_{\uparrow\downarrow}}n & gn
\end{bmatrix}.
\end{eqnarray}
\end{subequations}
The full elements of inverse Green's function at zero frequency (after analytic continuation) then becomes
\begin{equation}
G^{-1}_{\mathrm{ZM}}({\bf q},0)=-
\begin{bmatrix}
K^\mathrm{ZM}_0({\bf q})+\Sigma^\mathrm{ZM}_\mathrm{N} & \Sigma^\mathrm{ZM}_\mathrm{A}  
\\
\Sigma^\mathrm{ZM}_\mathrm{A} & K^\mathrm{ZM}_0(-{\bf q})+\Sigma^\mathrm{ZM}_\mathrm{N}
\end{bmatrix}.
\end{equation}
Therefore, we can straightforwardly find out the single-particle Green's function matrix $G_{\mathrm{B}}({\bf q},0)$ as well as the elements $G$ and $F$ defined in Eq.~\eqref{eq:G-and-F}. Thus, the new matrix $\mathcal{G}$ in the plane-wave and zero-momentum phases can be then obtained and substituted into Eq.~\eqref{eq:ns-Josephson} to calculate the superfluid density. We will verify the generalized Josephson relation at zero temperature by comparing the resulting superfluid density fraction with these obtained from other approaches in the following subsection (see Fig.~\ref{fig2}).

\subsection{Zero-temperature superfluid density from different approaches}
\begin{figure}[t]
\includegraphics[width=0.48\textwidth]{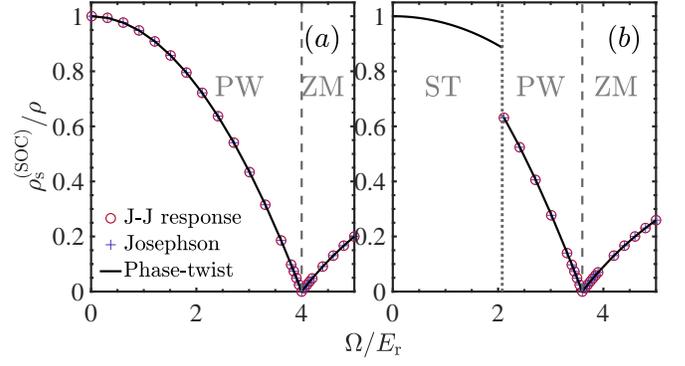}
\caption{Superfluid density fraction $\rho^{(\mathrm{SOC})}_{\mathrm{s}}/\rho$ in the SOC direction as a function of Rabi frequency at (a) equal and (b) unequal intra-, inter-spin interactions. Here the predictions given by the current-current response function and by the generalized Josephson relation are denoted by red circles and blue crosses, respectively. The black lines indicate the analytic prediction using a phase-twist approach in Ref.~\cite{chen2018quantum}.  The vertical dotted and dashed curves indicate the critical $\Omega_{c1}$ and $\Omega_{c2}$, respectively, i.e., Eqs.~\eqref{eq:omega1} and~\eqref{eq:omega2}. Here, we take the interaction energies [$G_1$, $G_2$]$/E_\mathrm{r}$= (a) [0.8, 0] and (b) [0.7, 0.1].}
\label{fig2}
\end{figure}

Here, we discuss the superfluid density in the SOC direction at zero temperature obtained from the current-current response function and from the generalized Josephson relation. In Fig.~\ref{fig2}, the superfluid density faction $\rho^{(\mathrm{SOC})}_{\mathrm{s}}/\rho$ in the uniform plane-wave and zero-momentum phase regimes is presented as a function of Rabi frequency at equal and unequal intra-, inter-spin interaction strengths.

It is clearly seen that the superfluid density $\rho^{(\mathrm{SOC})}_{\mathrm{s}}/\rho$ along the SOC direction calculated via Eq.~\eqref{eq:ns-JJ-response} from the current-current response function is identical to the one obtained via Eq.~\eqref{eq:ns-Josephson} from the generalized Josephson relation (see the red circles and blue crosses, respectively). Meanwhile, they agree with our previous analytic prediction in Eqs. (30-31) of Ref.~\cite{chen2018quantum} found using a phase-twist approach, i.e., $\rho^{(x,\mathrm{PW})}_{\mathrm{s}}/\rho = 1-\frac{E_{\mathrm{r}}}{(E_{\mathrm{r}}-G_{2})\Omega_{c2}^{2}/\Omega^{2}+G_{2}}$ and $\rho^{(x,\mathrm{ZM})}_{\mathrm{s}}/\rho = 1-\frac{4E_{\mathrm{r}}}{\Omega+4G_{2}}$, as shown in solid black lines.

The superfluid density exhibits a non-monotonic behavior with respect to Rabi frequency, decreasing smoothly in the PW phase towards zero at $\Omega_{c2}$ (vertical dashed line), and then rising back in the ZM phase. This non-trivial behavior at the PW-ZM phase transition point $\Omega_{c2}$ could be understood from its diverging effective mass, which suppresses the super-flow in the system~\cite{zheng2013properties}, and from the vanishing sound velocity or critical velocity favoring the creation of excitations to destroy its superfluidity~\cite{martone2012anisotropic,chen2017quantum,chen2018quantum}. It is worth noting that in Fig.~\ref{fig2}(b), the superfluid density in the stripe phase at $\Omega<\Omega_{c1}$ is merely calculated from the phase-twist approach in Ref.~\cite{chen2018quantum} and there's a discontinuity at $\Omega_{c1}$ due to the first-order nature of the ST-PW transition. In contrast, the critical Rabi frequency $\Omega_{c1}$ in Fig.~\ref{fig2}(a) shrinks to zero owing to the vanishing $G_2$.

Finally, we note that the calculation of the current-current response function and the generalization of the Josephson relation in the stripe phase would be much more involved. We will consider these two interesting issues in future works.

\subsection{Superfluid density, sound velocity and Landau critical velocity at finite temperature}
\begin{figure}[t]
\includegraphics[width=0.48\textwidth]{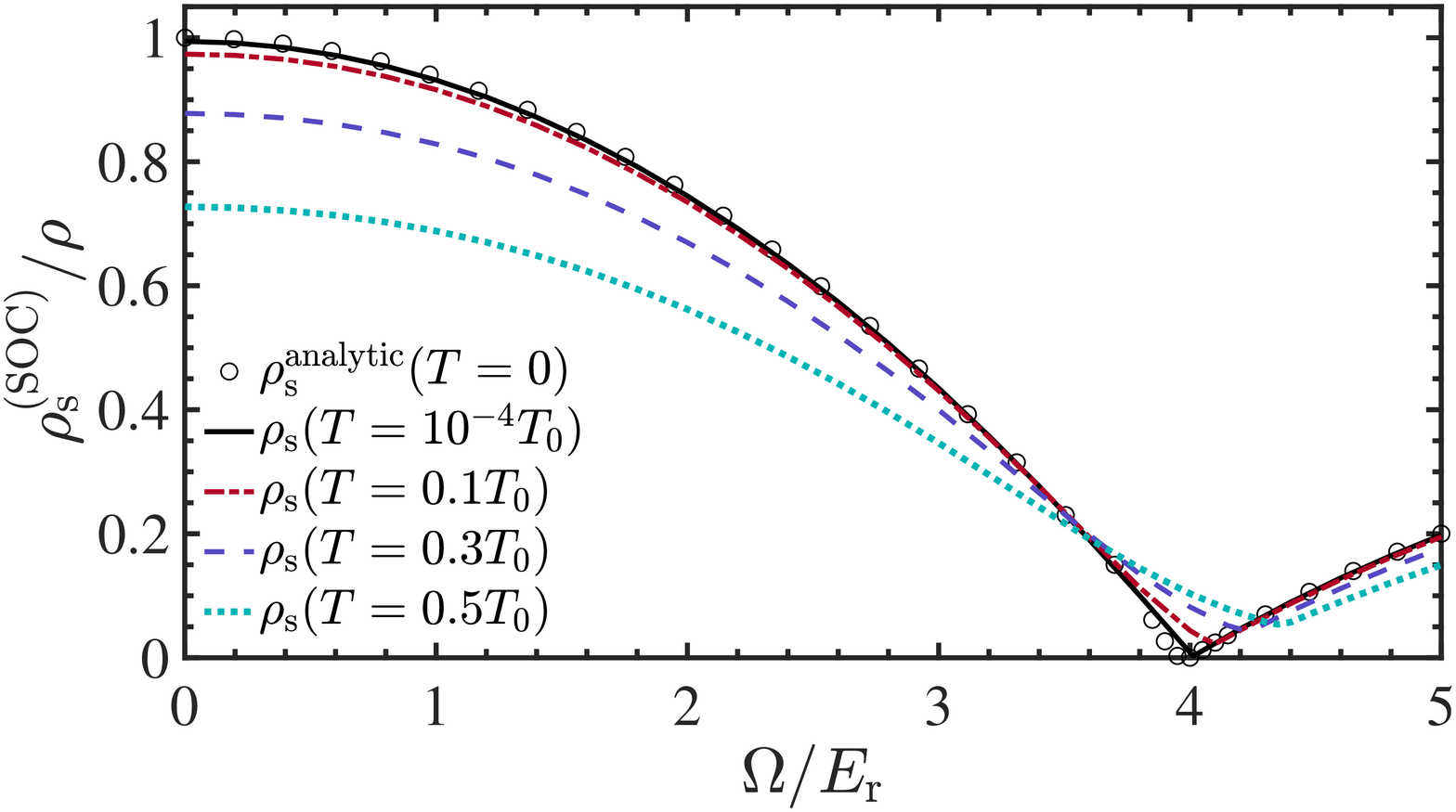}
\caption{Superfluid density fraction $\rho^{(\mathrm{SOC})}_{\mathrm{s}}/\rho$ in the SOC direction as a function of Rabi frequency at various values of temperature $T/T_0=10^{-4}$, $0.1$, $0.3$, and $0.5$ (i.e., lines) for a SOC $^{87}$Rb gas. Here $T_0$ is the critical BEC temperature of an ideal spinless Bose gas with density $n$, i.e., $T_{0}=2\pi\hbar^{2}[n/\zeta(3/2)]^{2/3}/(mk_{B})$, and the circles denote the zero-temperature analytic result from Ref.~\cite{chen2018quantum}. Other parameters are the same as in Fig.~\ref{fig1}.}
\label{fig3}
\end{figure}

We now turn to discuss the superfluid density at finite temperature using merely the current-current response function described in Sec.~\ref{sec:JJ-response-function}, and address its relation to sound velocity and Landau critical velocity.

In Fig.~\ref{fig3}, we present the behaviour of superfluid density $\rho_\mathrm{s}^\mathrm{(SOC)}$ in the SOC direction as a function of Rabi frequency in a SOC $^{87}$Rb Bose gas,  at four typical temperatures $T/T_0=10^{-4}$, $0.1$, $0.3$, and $0.5$ (i.e., lines) with $T_0$ being the critical BEC temperature of an ideal spinless Bose gas. We find that the superfluid density at a tiny temperature $T=10^{-4}T_0$ in solid black line is slightly different from the zero-temperature one denoted by the circles near the PW-ZM transition point, i.e., $\Omega_{c2}=4E_\mathrm{r}$. This is due to the quantum fluctuations taken into account in the current calculation, which can slightly shift the critical point and the minimum of sound velocity as discussed in Ref.~\cite{chen2017quantum}.

\begin{figure*}[ht]
\includegraphics[width=0.96\textwidth]{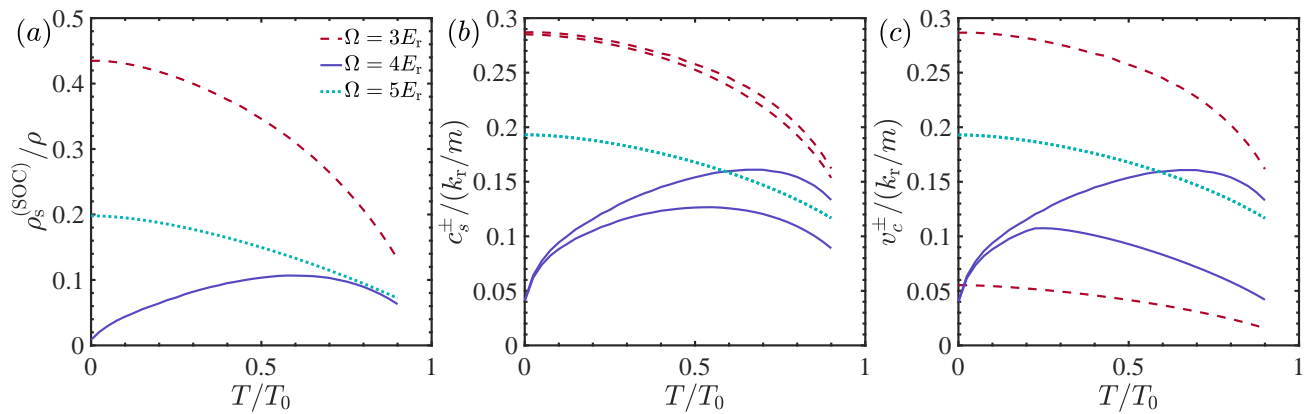}
\caption{(a) Superfluid density fraction $\rho^{(\mathrm{SOC})}_{\mathrm{s}}/\rho$ in the SOC direction, (b) sound velocities $c_s^{\pm}$ and (c) critical velocities $v_c^{\pm}$ in the $\pm x$ directions, as a function of temperature at three sets of Rabi frequency $\Omega/E_\mathrm{r}=3$, $4$, and $5$ for a SOC $^{87}$Rb gas. Other parameters are the same as in Fig.~\ref{fig1}.}
\label{fig4}
\end{figure*}

At zero temperature SOC plays a significant role on the superfluid density, leading to a non-monotonic behaviour with respect to Rabi frequency~\cite{zhang2016superfluid,chen2018quantum} as we already discussed in the last subsection. At nonzero temperature, thermal fluctuations start to play a non-negligible role, by suppressing the superfluidity and exhausting the condensate in the system. As a result, generally the superfluid density fraction reduces correspondingly with increasing temperature, as shown by the dashed, dotted, dashed-dotted colorful lines in Fig.~\ref{fig3}. In particular, as shown by the dotted and dashed lines in Fig.~\ref{fig4}(a), the superfluid density for a wide range of Rabi frequency exhibits a monotonic decreasing dependence on temperature.

However, near the PW-ZM transition point at $\Omega_{c2}=4E_\mathrm{r}$, this consensus is no longer true. There is a striking non-monotonic behavior of the superfluid density with respect to temperature, as shown by the solid blue line in Fig.~\ref{fig4}(a). As the temperature increases, the superfluid fraction increases first from a tiny value to the maximum about 0.1, and then decreases.

We attribute this interesting non-monotonic temperature dependence to the non-trivial interplay between the spin-orbit coupling and thermal fluctuations. At zero temperature, at the critical Rabi frequency the SOC effect strongly suppresses the superfluidity of the system with a vanishing sound velocity~\cite{zhang2012collective,zhang2016superfluid,chen2018quantum}. While at finite temperature, thermal fluctuations shift the critical point to larger Rabi frequency favoring the PW phase. As a result, we anticipate that the sound velocity may become nonzero, which instead restores the superfluidity of the system near the critical point~\cite{ji2015softening,chen2017quantum}. The anticipated temperature dependence of the sound velocity is examined in Fig.~\ref{fig4}(b). Indeed, near the transition point we find the very similar non-monotonic dependence as in the superfluid fraction. In contrast, away from the transition point in the deep PW and ZM phases (see, i.e., the dashed red and dotted blue lines for $\Omega/E_\mathrm{r}=3$ and $5$, respectively), the monotonic decreasing dependence on temperature is recovered in the sound velocity. This is consistent with the monotonic temperature dependence of the superfluid fraction at the corresponding Rabi frequency.

It is worth mentioning that, the unexpected, counterintuitive effect of thermal fluctuations near the PW-ZM transition point can persist up to a relatively large temperature and is responsible to the maximum superfluid fraction at $T\sim0.6T_0$.  As the temperature increases further, thermal fluctuations eventually depletes the superfluidity of the system. The superfluid fraction then decreases until reaching the BEC transition temperature.

We note also that, in the PW phase we have two sound velocities along and opposite to the SOC direction, owing to the emergence of the roton-maxon structure in the elementary excitation spectrum~\cite{martone2012anisotropic}. The non-trivial roton excitation spectrum also gives rise to two distinct Landau critical velocities $v_c^{\pm}$ in a PW phase~\cite{yu2017landau}, as shown in Fig.~\ref{fig4}(c), which provides additional information on the superfluidity. It is readily seen that, in the deep PW regime ($\Omega/E_\mathrm{r}=3$), one of the critical velocities in the dashed red line shows a significant difference with respect to the corresponding sound velocity due to the pronounced roton structure and a softened energy gap~\cite{chen2017quantum}. Near the transition point ($\Omega/E_\mathrm{r}=4$), the critical velocity in the solid blue line starts to deviate from the sound velocity at certain value of temperature $T/T_0\sim 0.2$. This deviation becomes even larger as the temperature rises, revealing that thermal fluctuations favor the emergence of the roton structure near the transition point.  The overall temperature dependence of Landau critical velocities are consistent with that of the superfluid fraction.

\section{Conclusions and Outlooks\label{sec:summary}}

In conclusions, we have studied the current-current response functions and the Josephson relation in a three-dimensional weakly interacting Bose gas with one-dimensional Raman-type spin-orbit coupling. The analytic expression of the current-current response functions and single-particle Green's functions are derived explicitly within the Bogoliubov approximation. The superfluid density fraction along the spin-orbit coupling direction is calculated using both approaches at zero temperature. The resulting superfluid density fraction agrees with previous analytic prediction obtained from a phase-twist approach. The zero-temperature superfluid density exhibits an intriguing behaviour as Rabi frequency rises, decreasing smoothly in the plane-wave phase, becoming zero at the critical point, and then rising back again in the zero-momentum phase.  

At finite temperature, we have calculated the superfluid density by using the current-current response functions, and have discussed its relation to sound velocity and Landau critical velocity. A significant non-monotonic temperature dependence is seen in these quantities near the transition point from the plane-wave to zero-momentum phases. Future works remain to be done, in order to calculate the current-current response functions and to establish the Josephson relation in the more interesting stripe phase.

\begin{acknowledgments}
X.L.C. acknowledges the insightful discussions with Zeng-Qiang Yu and the support from Swinburne University of Technology through an HDR Publication Award. Our research was supported by the Science Foundation of Zhejiang Sci-Tech University (ZSTU), Grant No. 21062339-Y (X.L.C.), and China Postdoctoral Science Foundation, Grant No. 2020M680495 (X.L.C.), the Australian Research Council's (ARC) Discovery Project, Grant No. DP180102018 (X.J.L.).
\end{acknowledgments}

\appendix
\begin{widetext}
\section{The current-current response functions at finite temperature\label{app:j-j-response}}

In this Appendix, we give an explicit derivation of the current-current response functions and other results in Sec.~\ref{sec:JJ-response-function}, in terms of the condensate wave function and the Bogoliubov wave functions at finite temperature.

By decoupling the Bose field operator $\hat{\Psi}_{\sigma}=\psi_{\sigma}+\hat{\eta}_{\sigma}$ in Eq.~\eqref{eq:newBosefield}, we first expand the spin-component current operators in Eq.~\eqref{eq:J-density-operator} for spin component $\sigma$ in the presence of the Raman-type spin-orbit coupling as $\hat{j}_{\sigma\sigma}({\bf r},t)=j_{0\sigma}({\bf r})+\delta\hat{j}_{\sigma}({\bf r},t)+\tilde{j}_{\sigma}({\bf r},t)$, where
\begin{subequations}
\begin{eqnarray}
j_{0\sigma}({\bf r}) &=& \frac{1}{2mi}(\psi_{\sigma}^*\nabla_x\psi_{\sigma}-\nabla_x\psi_{\sigma}^*\psi_{\sigma})\mp\frac{k_\mathrm{r}}{m}\psi_{\sigma}^*\psi_{\sigma}
\\
\delta\hat{j}_{\sigma}({\bf r},t) &=& \frac{1}{2mi}(\psi_{\sigma}^*\nabla_x\hat{\eta}_{\sigma}+\nabla_x\psi_{\sigma}\hat{\eta}_{\sigma}^*-\nabla_x\psi_{\sigma}^*\hat{\eta}_{\sigma}-\psi_{\sigma}\nabla_x\hat{\eta}_{\sigma}^*)\mp\frac{k_\mathrm{r}}{m}(\psi_{\sigma}^*\hat{\eta}_{\sigma}+\hat{\eta}_{\sigma}^*\psi_{\sigma})
\\
\tilde{j}_{\sigma}({\bf r},t) &=& \frac{1}{2mi}(\hat{\eta}_{\sigma}^*\nabla_x\hat{\eta}_{\sigma}-\nabla_x\hat{\eta}_{\sigma}^*\hat{\eta}_{\sigma})\mp\frac{k_\mathrm{r}}{m}\hat{\eta}_{\sigma}^*\hat{\eta}_{\sigma},
\end{eqnarray}
\end{subequations}
with the sign '$\mp$' is $-$ ($+$) for spin component $\sigma=\uparrow$ ($\downarrow$).

Within the Bogoliubov approximation, we are interested in the current fluctuation (i.e., $\delta \hat{j}$) only and the time-ordered correlation function for the current fluctuation density operator is given by
\begin{eqnarray}  \label{eq:jj-correlation}
\langle T_t\delta\hat{j}_{\sigma}({\bf r},t)\delta\hat{j}_{\sigma^\prime}({\bf r^\prime})\rangle &=&
\frac{1}{m^2}
\begin{pmatrix}\psi^*_{\sigma}({\bf r}) &\psi_{\sigma}({\bf r}) &\nabla_x\psi^*_{\sigma}({\bf r}) &\nabla_x\psi_{\sigma}({\bf r})\end{pmatrix}\begin{pmatrix}\mathcal{A} &\mathcal{B}\\ \mathcal{C} &\mathcal{D}\end{pmatrix}\begin{pmatrix}\psi^*_{\sigma^\prime}({\bf r^\prime})\\ \psi_{\sigma^\prime}({\bf r^\prime})\\ \nabla_x\psi^*_{\sigma^\prime}({\bf r^\prime})\\ \nabla_x\psi_{\sigma^\prime}({\bf r^\prime})\end{pmatrix},
\end{eqnarray} 
where we have introduced $\mathrm{sgn}(\sigma)=-1$ (1) for $\sigma=\uparrow$ ($\downarrow$), and the matrices read
\begin{subequations} \label{eq:matrix-ABCD}
\begin{eqnarray}
    \mathcal{A}&=&\frac{1}{4}\begin{pmatrix}
    -\langle \nabla_x\hat{\eta}_{\sigma}\nabla_x\hat{\eta}_{\sigma^\prime}\rangle
    &\langle \nabla_x\hat{\eta}_{\sigma}\nabla_x\hat{\eta}^*_{\sigma^\prime}\rangle 
    \\ 
    \langle \nabla_x\hat{\eta}^*_{\sigma}\nabla_x\hat{\eta}_{\sigma^\prime}\rangle &-\langle \nabla_x\hat{\eta}^*_{\sigma}\nabla_x\hat{\eta}^*_{\sigma^\prime}\rangle
    \end{pmatrix}+\frac{\mathrm{sgn}(\sigma^\prime)k_\mathrm{r}}{2i}\begin{pmatrix}
    \langle \nabla_x\hat{\eta}_{\sigma}\hat{\eta}_{\sigma^\prime}\rangle
    &\langle \nabla_x\hat{\eta}_{\sigma}\hat{\eta}^*_{\sigma^\prime}\rangle 
    \\ 
    -\langle \nabla_x\hat{\eta}^*_{\sigma}\hat{\eta}_{\sigma^\prime}\rangle &-\langle \nabla_x\hat{\eta}^*_{\sigma}\hat{\eta}^*_{\sigma^\prime}\rangle 
    \end{pmatrix}\nonumber
    \\
    &&+\frac{\mathrm{sgn}(\sigma)k_\mathrm{r}}{2i}\begin{pmatrix}
    \langle \hat{\eta}_{\sigma}\nabla_x\hat{\eta}_{\sigma^\prime}\rangle 
    &-\langle \hat{\eta}_{\sigma}\nabla_x\hat{\eta}^*_{\sigma^\prime}\rangle \\ 
    \langle \hat{\eta}^*_{\sigma}\nabla_x\hat{\eta}_{\sigma^\prime}\rangle &-\langle \hat{\eta}^*_{\sigma}\nabla_x\hat{\eta}^*_{\sigma^\prime}\rangle
    \end{pmatrix}
    +\mathrm{sgn}(\sigma)\mathrm{sgn}(\sigma^\prime)k_\mathrm{r}^2\begin{pmatrix}
    \langle \hat{\eta}_{\sigma}\hat{\eta}_{\sigma^\prime}\rangle 
    &\langle \hat{\eta}_{\sigma}\hat{\eta}^*_{\sigma^\prime}\rangle \\ 
    \langle \hat{\eta}^*_{\sigma}\hat{\eta}_{\sigma^\prime}\rangle &\langle \hat{\eta}^*_{\sigma}\hat{\eta}^*_{\sigma^\prime}\rangle
    \end{pmatrix},
\\
     \mathcal{B}&=&\frac{1}{4}\begin{pmatrix}
    \langle \nabla_x\hat{\eta}_{\sigma}\hat{\eta}_{\sigma^\prime}\rangle &-\langle \nabla_x\hat{\eta}_{\sigma}\hat{\eta}^*_{\sigma^\prime}\rangle
    \\ 
    -\langle \nabla_x\hat{\eta}^*_{\sigma}\hat{\eta}_{\sigma^\prime}\rangle  &\langle \nabla_x\hat{\eta}^*_{\sigma}\hat{\eta}^*_{\sigma^\prime}\rangle
    \end{pmatrix}
    +\frac{\mathrm{sgn}(\sigma)k_\mathrm{r}}{2i}\begin{pmatrix}
    -\langle \hat{\eta}_{\sigma}\hat{\eta}_{\sigma^\prime}\rangle  &\langle \hat{\eta}_{\sigma}\hat{\eta}^*_{\sigma^\prime}\rangle
    \\ 
    -\langle \hat{\eta}^*_{\sigma}\hat{\eta}_{\sigma^\prime}\rangle  &\langle \hat{\eta}^*_{\sigma}\hat{\eta}^*_{\sigma^\prime}\rangle
    \end{pmatrix},
\\
    \mathcal{C}&=&\frac{1}{4}\begin{pmatrix}
    \langle \hat{\eta}_{\sigma}\nabla_x\hat{\eta}_{\sigma^\prime}\rangle &-\langle \hat{\eta}_{\sigma}\nabla_x\hat{\eta}^*_{\sigma^\prime}\rangle
    \\ 
    -\langle \hat{\eta}^*_{\sigma}\nabla_x\hat{\eta}_{\sigma^\prime}\rangle &\langle \hat{\eta}^*_{\sigma}\nabla_x\hat{\eta}^*_{\sigma^\prime}\rangle 
    \end{pmatrix}
    +\frac{\mathrm{sgn}(\sigma^\prime)k_\mathrm{r}}{2i}\begin{pmatrix}
    -\langle \hat{\eta}_{\sigma}\hat{\eta}_{\sigma^\prime}\rangle &-\langle \hat{\eta}_{\sigma}\hat{\eta}^*_{\sigma^\prime}\rangle
    \\ 
    \langle \hat{\eta}^*_{\sigma}\hat{\eta}_{\sigma^\prime}\rangle &\langle \hat{\eta}^*_{\sigma}\hat{\eta}^*_{\sigma^\prime}\rangle 
    \end{pmatrix},
\\    
    \mathcal{D}&=&\frac{1}{4}\begin{pmatrix}
    -\langle \hat{\eta}_{\sigma}\hat{\eta}_{\sigma^\prime}\rangle &\langle \hat{\eta}_{\sigma}\hat{\eta}^*_{\sigma^\prime}\rangle
    \\ 
    \langle \hat{\eta}^*_{\sigma}\hat{\eta}_{\sigma^\prime}\rangle &-\langle \hat{\eta}^*_{\sigma}\hat{\eta}^*_{\sigma^\prime}\rangle
    \end{pmatrix}.
\end{eqnarray}
\end{subequations}
Here $\langle \cdots\rangle$ is the thermal average in the statistical equilibrium and $\hat{\eta}_{\sigma}\hat{\eta}_{\sigma^\prime}\equiv\hat{\eta}_{\sigma}({\bf r},t)\hat{\eta}_{\sigma^\prime}({\bf r^\prime})$. After taking a plane-wave {\it Ansatz} $\psi_{\sigma}({\bf r})=\psi_{\sigma}e^{iP_xx}$ for the condensate wave function, we can further rewrite the fluctuation correlation function in the imaginary time ($\tau=it$) as
\begin{eqnarray} \label{eq:j-jprime}
\langle T_{\tau}\delta\hat{j}_{\sigma}({\bf r},\tau)\delta\hat{j}_{\sigma^\prime}({\bf r^\prime})\rangle &=&
\mathcal{A}_1+\mathcal{A}_2+\mathcal{A}_3+\mathcal{A}_4
\end{eqnarray} 
with
\begin{subequations} \label{eq:A-matrices}
\begin{eqnarray}
\mathcal{A}_1&=&\frac{1}{4m^2}\begin{pmatrix}\psi^*_{\sigma}({\bf r}) &\psi_{\sigma}({\bf r}) \end{pmatrix}\begin{pmatrix}
    -\langle \nabla_x\hat{\eta}_{\sigma}\nabla_x\hat{\eta}_{\sigma^\prime}\rangle
    &\langle \nabla_x\hat{\eta}_{\sigma}\nabla_x\hat{\eta}^*_{\sigma^\prime}\rangle 
    \\ 
    \langle \nabla_x\hat{\eta}^*_{\sigma}\nabla_x\hat{\eta}_{\sigma^\prime}\rangle &-\langle \nabla_x\hat{\eta}^*_{\sigma}\nabla_x\hat{\eta}^*_{\sigma^\prime}\rangle
    \end{pmatrix}\begin{pmatrix}\psi^*_{\sigma^\prime}({\bf r^\prime})\\ \psi_{\sigma^\prime}({\bf r^\prime})\end{pmatrix},
    \\
\mathcal{A}_2&=&\frac{P_x+2bk_\mathrm{r}}{4m^2i}\begin{pmatrix}\psi^*_{\sigma}({\bf r}) &\psi_{\sigma}({\bf r}) \end{pmatrix}\begin{pmatrix}
    \langle \nabla_x\hat{\eta}_{\sigma}\hat{\eta}_{\sigma^\prime}\rangle
    &\langle \nabla_x\hat{\eta}_{\sigma}\hat{\eta}^*_{\sigma^\prime}\rangle 
    \\ 
    -\langle \nabla_x\hat{\eta}^*_{\sigma}\hat{\eta}_{\sigma^\prime}\rangle &-\langle \nabla_x\hat{\eta}^*_{\sigma}\hat{\eta}^*_{\sigma^\prime}\rangle 
    \end{pmatrix}\begin{pmatrix}\psi^*_{\sigma^\prime}({\bf r^\prime})\\ \psi_{\sigma^\prime}({\bf r^\prime})\end{pmatrix},
\\
\mathcal{A}_3&=&\frac{P_x+2ak_\mathrm{r}}{4m^2i}\begin{pmatrix}\psi^*_{\sigma}({\bf r}) &\psi_{\sigma}({\bf r}) \end{pmatrix}\begin{pmatrix}
    \langle \hat{\eta}_{\sigma}\nabla_x\hat{\eta}_{\sigma^\prime}\rangle 
    &-\langle \hat{\eta}_{\sigma}\nabla_x\hat{\eta}^*_{\sigma^\prime}\rangle \\ 
    \langle \hat{\eta}^*_{\sigma}\nabla_x\hat{\eta}_{\sigma^\prime}\rangle &-\langle \hat{\eta}^*_{\sigma}\nabla_x\hat{\eta}^*_{\sigma^\prime}\rangle
    \end{pmatrix}\begin{pmatrix}\psi^*_{\sigma^\prime}({\bf r^\prime})\\ \psi_{\sigma^\prime}({\bf r^\prime})\end{pmatrix},
\\
\mathcal{A}_4&=&\frac{1}{m^2}\left[abk_\mathrm{r}^2+\frac{k_\mathrm{r}P_x}{2}(a+b)+\frac{P_x^2}{4}\right]\begin{pmatrix}\psi^*_{\sigma}({\bf r}) &\psi_{\sigma}({\bf r}) \end{pmatrix}\begin{pmatrix}
    \langle \hat{\eta}_{\sigma}\hat{\eta}_{\sigma^\prime}\rangle &\langle \hat{\eta}_{\sigma}\hat{\eta}^*_{\sigma^\prime}\rangle
    \\ 
    \langle \hat{\eta}^*_{\sigma}\hat{\eta}_{\sigma^\prime}\rangle &\langle \hat{\eta}^*_{\sigma}\hat{\eta}^*_{\sigma^\prime}\rangle
    \end{pmatrix}\begin{pmatrix}\psi^*_{\sigma^\prime}({\bf r^\prime})\\ \psi_{\sigma^\prime}({\bf r^\prime})\end{pmatrix}
\end{eqnarray}
\end{subequations}
and the coefficients $a\equiv\mathrm{sgn}(\sigma)$, $b\equiv\mathrm{sgn}(\sigma^\prime)$.

At finite temperature $\beta\equiv 1/k_BT$, the current response function is then calculated collectively by using Eq.~\eqref{eq:delta-J} as
\begin{eqnarray}  
&&\chi_{JJ}({\bf r},{\bf r^\prime};i\omega_n)\equiv\chi^{\uparrow\uparrow}+\chi^{\uparrow\downarrow}+\chi^{\downarrow\uparrow}+\chi^{\downarrow\downarrow}=\int_0^\beta d\tau e^{i\omega_n\tau} \frac{\theta(\tau)}{V}\langle T_{\tau}\delta\hat{j}({\bf r},\tau)\delta\hat{j}({\bf r^\prime})\rangle.
\end{eqnarray}
We can take an example for illustration of calculating $\langle T_{\tau}\delta\hat{j}_{\sigma}({\bf r},\tau)\delta\hat{j}_{\sigma^\prime}({\bf r^\prime})\rangle$ in Eq.~\eqref{eq:j-jprime}. For instance,
\begin{eqnarray}  
&&\mathcal{A}_1^{11}({\bf r},{\bf r^\prime};\tau)=\frac{1}{4m^2}\psi^*_{\sigma}({\bf r})(-)\langle \nabla_x\hat{\eta}_{\sigma}({\bf r},\tau)\nabla_x\hat{\eta}_{\sigma^\prime}({\bf r^\prime})\rangle\psi^*_{\sigma^\prime}({\bf r^\prime}) \nonumber
\\
&=&\frac{1}{4m^2}\psi^*_{\sigma}({\bf r})(-)\psi^*_{\sigma^\prime}({\bf r^\prime})\sum_{j\equiv({\bf q}_1,\lambda_1),k\equiv({\bf q}_2,\lambda_2)} \left\langle\nabla_x\left[e^{iP_xx}\left(u^{(\lambda_1)}_{{\bf q}_1\sigma}e^{i{\bf q_1r}}\hat{\alpha}_je^{-\omega_j \tau}+(v^{(\lambda_1)}_{{\bf q}_1\sigma})^*e^{-i{\bf q_1r}}\hat{\alpha}^\dagger_je^{\omega_j \tau}\right)\right]\right. \nonumber
\\
&& \left.\nabla_x\left[e^{iP_xx}\left(u^{(\lambda_2)}_{{\bf q}_2\sigma^\prime}e^{i{\bf q}_2{\bf r^\prime}}\hat{\alpha}_k+(v^{(\lambda_2)}_{{\bf q}_2\sigma^\prime})^*e^{-i{\bf q}_2{\bf r^\prime}}\hat{\alpha}^\dagger_k\right)\right]\right\rangle \nonumber
\\
&=& \frac{\psi_{\sigma}\psi_{\sigma^\prime}}{4m^2}e^{i{\bf q(r-r^\prime)}}\sum_{j}(P_x^2-q_x^2)\left[u_{\sigma}v_{\sigma^\prime}\left(1+f_B(\omega_j)\right)e^{-\omega_j \tau} +v_{\sigma}u_{\sigma^\prime}f_\mathrm{B}(\omega_j)e^{\omega_j \tau}\right],
\end{eqnarray} 
with the expanded fluctuation operator $\hat{\eta}_{\sigma}({\bf r},t)=e^{iP_xx}\sum_{j\equiv({\bf q},\lambda)}\left(u^{(\lambda)}_{{\bf q}\sigma}e^{i{\bf qr}}\hat{\alpha}_je^{-i\omega_j t}+(v^{(\lambda)}_{{\bf q}\sigma})^*e^{-i{\bf qr}}\hat{\alpha}^\dagger_je^{i\omega_j t}\right)$ in a quasiparticle basis and $u_{\sigma}\equiv u^{(\lambda)}_{{\bf q}\sigma}$ and $v_{\sigma}\equiv v^{(\lambda)}_{{\bf q}\sigma}$ defined here and hereafter. $f_\mathrm{B}(\omega_j)\equiv\langle\hat{\alpha}^\dagger_j\hat{\alpha}_j\rangle$ is the occupation number for $j$-th quasiparticle with energy $\omega_j$ satisfying the bosonic statistical distribution function as $f_\mathrm{B}(\omega_j)=1/(\mathrm{exp}(\beta\omega_j)-1)$.

After integrating $\int_0^\beta e^{(i\omega_n-\omega)\tau}\theta(\tau) d\tau =\frac{e^{\beta(i\omega_n-\omega)}-1}{i\omega_n-\omega}$ with the Matsubara frequencies $i\omega_n\equiv2n\pi i/\beta$ for bosons and $e^{\beta i\omega_n}=1$, the response function becomes
\begin{eqnarray}  
\chi^{11}_{1}({\bf r},{\bf r^\prime}; i\omega_n) &=& \int_0^\beta d\tau e^{i\omega_n\tau}\frac{\theta(\tau)}{V}  \mathcal{A}_1^{11}({\bf r},{\bf r^\prime};\tau) \nonumber
\\
&=& \frac{\psi_{\sigma}\psi_{\sigma^\prime}}{4m^2V}e^{i{\bf q(r-r^\prime)}}\sum_{j}(P_x^2-q_x^2)\int_0^\beta d\tau \theta(\tau)\left[u_{\sigma}v_{\sigma^\prime}\left(1+f_B(\omega_j)\right)e^{(i\omega_n-\omega_j) \tau} +v_{\sigma}u_{\sigma^\prime}f_\mathrm{B}(\omega_j)e^{(i\omega_n+\omega_j) \tau}\right] \nonumber
\\
&=& \frac{\psi_{\sigma}\psi_{\sigma^\prime}}{4m^2V}e^{i{\bf q(r-r^\prime)}}\sum_{j}(P_x^2-q_x^2)\left[\frac{v_{\sigma}u_{\sigma^\prime}}{i\omega_n+\omega_j}-\frac{u_{\sigma}v_{\sigma^\prime}}{i\omega_n-\omega_j}\right],
\end{eqnarray}
where we have used the relations $(e^{-\beta\omega_j}-1)\left(1+f_B(\omega_j)\right)=-1$ and $(e^{\beta\omega_j}-1)f_\mathrm{B}(\omega_j)=1$. The real-time response function can be then obtained by employing the analytic continuation $(i\omega_n\rightarrow\omega+i\eta)$ with $\eta=0^+$ as
\begin{equation}
\chi^{11}_{1}({\bf r},{\bf r^\prime};\omega+i\eta)=\frac{\psi_{\sigma}\psi_{\sigma^\prime}}{4m^2V}e^{i{\bf q(r-r^\prime)}}\sum_{j}(P_x^2-q_x^2)\left[\frac{v_{\sigma}u_{\sigma^\prime}}{\omega+i\eta+\omega_j}-\frac{u_{\sigma}v_{\sigma^\prime}}{\omega+i\eta-\omega_j}\right].
\end{equation}
Similarly, another three response functions can be obtained as
\begin{subequations}
\begin{eqnarray} 
\chi^{12}_{1}({\bf r},{\bf r^\prime};\omega+i\eta)&=&\frac{\psi_{\sigma}\psi_{\sigma^\prime}}{4m^2V}e^{i{\bf q(r-r^\prime)}}\sum_{j}\left[(P_x-q_x)^2\frac{v_{\sigma}v_{\sigma^\prime}}{\omega+i\eta+\omega_j}-(P_x+q_x)^2\frac{u_{\sigma}u_{\sigma^\prime}}{\omega+i\eta-\omega_j}\right],\\
\chi^{21}_{1}({\bf r},{\bf r^\prime};\omega+i\eta)&=&\frac{\psi_{\sigma}\psi_{\sigma^\prime}}{4m^2V}e^{i{\bf q(r-r^\prime)}}\sum_{j}\left[(P_x+q_x)^2\frac{u_{\sigma}u_{\sigma^\prime}}{\omega+i\eta+\omega_j}-(P_x-q_x)^2\frac{v_{\sigma}v_{\sigma^\prime}}{\omega+i\eta-\omega_j}\right],\\
\chi^{22}_{1}({\bf r},{\bf r^\prime};\omega+i\eta)&=&\frac{\psi_{\sigma}\psi_{\sigma^\prime}}{4m^2V}e^{i{\bf q(r-r^\prime)}}\sum_{j}(P_x^2-q_x^2)\left[\frac{u_{\sigma}v_{\sigma^\prime}}{\omega+i\eta+\omega_j}-\frac{v_{\sigma}u_{\sigma^\prime}}{\omega+i\eta-\omega_j}\right].
\end{eqnarray}
\end{subequations}
The static response  for ${\bf q}$-component can be then obtained by setting $\omega=0$ and taking a Fourier transform to momentum space, and the equations now read
\begin{subequations}
\begin{eqnarray} 
\chi^{11}_{1}({\bf q};0)&=&\chi^{22}_{1}({\bf q};0)=\frac{\psi_{\sigma}\psi_{\sigma^\prime}}{4m^2V}\sum_{j}(P_x^2-q_x^2)\left[\frac{u_{\sigma}v_{\sigma^\prime}+v_{\sigma}u_{\sigma^\prime}}{\omega_j}\right],
\\
\chi^{12}_{1}({\bf q};0)&=&\chi^{21}_{1}({\bf q};0)=\frac{\psi_{\sigma}\psi_{\sigma^\prime}}{4m^2V}\sum_{j}\left[\frac{(P_x+q_x)^2u_{\sigma}u_{\sigma^\prime}+(P_x-q_x)^2v_{\sigma}v_{\sigma^\prime}}{\omega_j}\right].
\end{eqnarray}
\end{subequations}

Similarly, the remaining static response functions for $\mathcal{A}_{2,3,4}$ can be obtained as
\begin{subequations}
\begin{eqnarray} 
\chi^{11}_{2}({\bf q};0)&=&\chi^{22}_{2}({\bf q};0)=\frac{\psi_{\sigma}\psi_{\sigma^\prime}}{4m^2V}\sum_{j}(P_x+2bk_\mathrm{r})\left[\frac{(P_x+q_x)u_{\sigma}v_{\sigma^\prime}+(P_x-q_x)v_{\sigma}u_{\sigma^\prime}}{\omega_j}\right],
\\
\chi^{12}_{2}({\bf q};0)&=&\chi^{21}_{2}({\bf q};0)=\frac{\psi_{\sigma}\psi_{\sigma^\prime}}{4m^2V}\sum_{j}(P_x+2bk_\mathrm{r})\left[\frac{(P_x+q_x)u_{\sigma}u_{\sigma^\prime}+(P_x-q_x)v_{\sigma}v_{\sigma^\prime}}{\omega_j}\right],
\\ 
\chi^{11}_{3}({\bf q};0)&=&\chi^{22}_{3}({\bf q};0)=\frac{\psi_{\sigma}\psi_{\sigma^\prime}}{4m^2V}\sum_{j}(P_x+2ak_\mathrm{r})\left[\frac{(P_x-q_x)u_{\sigma}v_{\sigma^\prime}+(P_x+q_x)v_{\sigma}u_{\sigma^\prime}}{\omega_j}\right],
\\
\chi^{12}_{3}({\bf q};0)&=&\chi^{21}_{3}({\bf q};0)=\frac{\psi_{\sigma}\psi_{\sigma^\prime}}{4m^2V}\sum_{j}(P_x+2ak_\mathrm{r})\left[\frac{(P_x+q_x)u_{\sigma}u_{\sigma^\prime}+(P_x-q_x)v_{\sigma}v_{\sigma^\prime}}{\omega_j}\right],
\\
\chi^{11}_{4}({\bf q};0)&=&\chi^{22}_{4}({\bf q};0)=\frac{\psi_{\sigma}\psi_{\sigma^\prime}}{m^2V}\sum_{j}\left[abk_\mathrm{r}^2+\frac{k_\mathrm{r}P_x}{2}(a+b)+\frac{P_x^2}{4}\right]\left[\frac{u_{\sigma}v_{\sigma^\prime}+v_{\sigma}u_{\sigma^\prime}}{\omega_j}\right],
\\
\chi^{12}_{4}({\bf q};0)&=&\chi^{21}_{4}({\bf q};0)=\frac{\psi_{\sigma}\psi_{\sigma^\prime}}{m^2V}\sum_{j}\left[abk_\mathrm{r}^2+\frac{k_\mathrm{r}P_x}{2}(a+b)+\frac{P_x^2}{4}\right]\left[\frac{u_{\sigma}u_{\sigma^\prime}+v_{\sigma}v_{\sigma^\prime}}{\omega_j}\right].
\end{eqnarray}
\end{subequations}

Therefore, the static response functions are derived by summarizing $\chi^{\sigma\sigma^\prime}({\bf q};0)=\sum_{m,\mu\nu}\left[\chi^{\mu\nu}_{m}({\bf q};0)\right]_{\sigma\sigma^\prime}$ for spin components $\sigma\sigma'$ as
\begin{eqnarray}
\chi^{11}_{\sigma\sigma^\prime}({\bf q};0)&=&\chi^{22}_{\sigma\sigma^\prime}({\bf q};0)
=\frac{\psi_{\sigma}\psi_{\sigma^\prime}}{4m^2V}\sum_{j} \frac{B_1u_{\sigma}v_{\sigma^\prime}+B_2v_{\sigma}u_{\sigma^\prime}}{\omega_j},
\\
\chi^{12}_{\sigma\sigma^\prime}({\bf q};0)&=&\chi^{21}_{\sigma\sigma^\prime}({\bf q};0) =\frac{\psi_{\sigma}\psi_{\sigma^\prime}}{4m^2V} \sum_{j} \frac{B_3u_{\sigma}u_{\sigma^\prime}+B_4v_{\sigma}v_{\sigma^\prime}}{\omega_j},
\end{eqnarray}
with $B_1 = (2ak_\mathrm{r}+2P_x+q_x)(2bk_\mathrm{r}+2P_x-q_x)$, $B_2 = (2ak_\mathrm{r}+2P_x-q_x)(2bk_\mathrm{r}+2P_x+q_x)$, $B_3 = (2ak_\mathrm{r}+2P_x+q_x)(2bk_\mathrm{r}+2P_x+q_x)$ and $B_4 = (2ak_\mathrm{r}+2P_x-q_x)(2bk_\mathrm{r}+2P_x-q_x)$. Thus, for spin index $\sigma\sigma'=(\uparrow\uparrow,\downarrow\downarrow,\uparrow\downarrow,\downarrow\uparrow)$, we obtain the explicit expressions for Eq.~\eqref{eq:ssprime-response} in the main text as
\begin{subequations}
\begin{eqnarray}
\chi^{\uparrow\uparrow}({\bf q};0)&=&\sum_{ij}\chi^{ij}_{\uparrow\uparrow} = \frac{\psi_{\uparrow}^2}{m^2V}\sum_{j}\frac{C_1u_{\uparrow}u_{\uparrow}+C_2v_{\uparrow}v_{\uparrow}+C_3u_{\uparrow}v_{\uparrow}}{\omega_j},
\\
\chi^{\downarrow\downarrow}({\bf q};0)&=&\sum_{ij}\chi^{ij}_{\downarrow\downarrow} = \frac{\psi_{\downarrow}^2}{m^2V}\sum_{j}\frac{C_4u_{\downarrow}u_{\downarrow}+C_5v_{\downarrow}v_{\downarrow}+C_6u_{\downarrow}v_{\downarrow}}{\omega_j},
\\
\chi^{\uparrow\downarrow}({\bf q};0)&=&\chi^{\downarrow\uparrow}({\bf q};0)=\sum_{ij}\chi^{ij}_{\uparrow\downarrow} =\frac{\psi_{\uparrow}\psi_{\downarrow}}{m^2V}\sum_{j}\frac{C_7u_{\uparrow}u_{\downarrow}+C_8v_{\uparrow}v_{\downarrow}+C_9u_{\uparrow}v_{\downarrow}+C_{10}v_{\uparrow}u_{\downarrow}}{\omega_j},
\end{eqnarray}
\end{subequations}
with
\begin{subequations}
\begin{eqnarray}
C_1 &=& \frac{1}{2}(2P_x+q_x-2k_\mathrm{r})^2, 
\quad C_2 = \frac{1}{2}(2P_x-q_x-2k_\mathrm{r})^2, 
\\
C_3 &=& (2P_x-q_x-2k_\mathrm{r})(2P_x+q_x-2k_\mathrm{r}),  
\\
C_4 &=& \frac{1}{2}(2P_x+q_x+2k_\mathrm{r})^2,  
\quad C_5 = \frac{1}{2}(2P_x-q_x+2k_\mathrm{r})^2,  
\\
C_6 &=& (2P_x-q_x+2k_\mathrm{r})(2P_x+q_x+2k_\mathrm{r}),  
\\
C_7 &=& \frac{1}{2}(2P_x+q_x)^2-2k_\mathrm{r}^2, 
\quad C_8 = \frac{1}{2}(2P_x-q_x)^2-2k_\mathrm{r}^2,  
\\
C_9 &=& 2P_x^2-\frac{1}{2}(q_x-2k_\mathrm{r})^2, 
\quad C_{10} = 2P_x^2-\frac{1}{2}(q_x+2k_\mathrm{r})^2.
\end{eqnarray}
\end{subequations}
In the zero-momentum phase with $P_x=0$, the coefficients $C_n$ reduces to $C_1 = C_5 = -C_9 = \frac{1}{2}(q_x-2k_\mathrm{r})^2$, $C_2 = C_4 = -C_{10} = \frac{1}{2}(q_x+2k_\mathrm{r})^2$ and $C_3 = C_6 = -2C_7 = -2C_8 = 4k_\mathrm{r}^2-q_x^2$.

Eventually, the static current-current response function, i.e., Eq.~\eqref{eq:chi_JJ_rr} in the main text, is given by
\begin{equation}
    \chi_{JJ}({\bf q}; 0)=\chi^{\uparrow\uparrow}({\bf q};0)+\chi^{\downarrow\downarrow}({\bf q};0)+\chi^{\uparrow\downarrow}({\bf q};0)+\chi^{\downarrow\uparrow}({\bf q};0).
\end{equation}

\end{widetext}

\bibliography{references}
\end{document}